\documentstyle[aps,prb,epsf]{revtex}
\begin{document}
\draft
\title{Current responsivity of semiconductor superlattice
THz-photon detectors}
\author{Anatoly A. Ignatov\cite{AA} and Antti-Pekka Jauho}
\address{Mikroelektronik Centret, Technical University of Denmark,
Building 345 east, DK-2800 Lyngby, Denmark}
\maketitle
\begin{abstract}
The  current responsivity of a semiconductor superlattice
THz-photon  detector  is calculated using  an  equivalent
circuit  model  which  takes  into  account  the   finite
matching  efficiency between a detector antenna  and  the
superlattice   in  the  presence  of  parasitic   losses.
Calculations    performed   for    currently    available
superlattice diodes show that both the magnitudes and the
roll-off  frequencies  of the responsivity  are  strongly
influenced   by  an  excitation  of  hybrid  plasma-Bloch
oscillations  which  are found to be  eigenmodes  of  the
system  in  the  THz-frequency band.  The  expected  room
temperature values of the responsivity ( 2--3 A/W in the 1--3 
THz-frequency band) range up to several percents of the
quantum  efficiency $e/\hbar\omega$  of  an ideal superconductor  tunnel
junction   detector.   Properly  designed   semiconductor
superlattice detectors may thus demonstrate  better  room
temperature  THz-photon  responsivity  than  conventional
Schottky junction devices.
\end{abstract}


\section{INTRODUCTION}
Theoretical   investigations   of   high-frequency
properties  of  semiconductor  superlattices  have   been
carried  out  for almost thirty years starting  from  the
trail-blazing works by Esaki and Tsu\cite{ref1,ref2}. They proposed to
prepare  superlattices  made of  different  semiconductor
materials  in  order to realise an artificial  periodical
system that would allow to observe Bloch oscillations due
to  Bragg reflection of electrons from the boundaries  of
the  Brillouin  zone.  Making use  of  the  semiclassical
Boltzmann equation they found that the dc current-voltage
characteristics  of  the  superlattice  should   show   a
negative differential conductance in sufficiently  strong
electric fields, namely, when the Bloch frequency
$\Omega_B = eEd/\hbar$  (where
$e$ is the electron's charge, $E$ is the dc electric field in
the  superlattice, $d$ is the superlattice period,  
$\hbar$ is  the
Plank's  constant) becomes comparable  to  the  effective
scattering frequency $\nu$\cite{ref1}. They proposed to make use of the
negative differential conductance for development  of  an
oscillator in the THz-frequency band\cite{ref1} and also  suggested
that  the  superlattice might posses promising properties
as  an  artificial  material  for  non-linear  mixing  of
photons\cite{ref2}.

    Later, several papers have proposed\cite{ref3,ref4,ref5} 
to probe Bloch
oscillations  in a superlattice in a strong  dc  electric
field  by  applying an additional strong ac field.  Self-
induced transparency\cite{ref3}, dc current suppression by a strong
ac  field causing dynamical localisation of carriers
\cite{ref3,ref4,ref5,ref6,ref7,ref8,ref9,ref10,ref11,ref12,ref13,ref14,ref15},
absolute  negative  conductance\cite{ref3}, and  current  harmonics
generation with oscillating power-dependence\cite{ref3,ref5} have been
predicted   to   occur  in  the  presence   of   resonant
interaction  between Bloch oscillating  electrons  and  a
strong THz-frequency radiation.

     A  profound analogy between the dynamics of electron
wave-packets  in  periodic systems and  the  dynamics  of
superconducting Josephson junctions\cite{ref16,ref17} was put  forward
to  explain  the  main features of the superlattice  THz-
frequency   response\cite{ref17}.  It  has   been   suggested,   in
particular,  that the resonant steps on the  dc  current-
voltage characteristics of the irradiated superlattices\cite{ref17}
should  occur  in  analogy with `Shapiro steps'  normally
observable  in  irradiated  superconducting  junctions\cite{ref18}.
Several recent papers show that the curious behaviour  of
electrons  in the superlattices may give rise to  a  rich
variety of new non-linear phenomena occurring in the THz-
field irradiated superlattices
\cite{ref19,ref20,ref21,ref22,ref23,ref24}

    In a number of papers Esaki-Tsu negative differential
conductance    was    found    to    be    experimentally
realisable\cite{ref25,ref26}  and has been carefully examined
\cite{ref27,ref28,ref29}  in
the  context  of development of new millimetre-wave  band
(0.03-0.3 THz) oscillators. On the other hand, the recent
observations   of   a   strong  dc  current   suppression
indicating   dynamical   localisation   of   electrons\cite{ref30},
absolute  negative conductance\cite{ref31,ref32}, and `Shapiro  steps'
on   the   dc  current-voltage  curve  of  the  THz-field
irradiated   superlattices\cite{ref33}  open  the   prospects   for
applications  of the superlattices as novel  solid  state
detectors operating in 1-10 THz-frequency band which  the
Bloch  frequency  in the superlattices  normally  belongs
to\cite{ref34}.

     It  has  recently  been estimated\cite{ref35}  that  the  room
temperature   current  responsivity  of  a   superlattice
detector  ideally coupled to the THz-photons  can  nearly
reach  the  quantum efficiency  $e/\hbar\omega$
(where $\omega$  is  the  incident
radiation  frequency) in the limit of high frequencies
$\omega\gg\nu$.
This   value  of  the  responsivity  is  being   normally
considered  as  a  quantum limit for detectors  based  on
superconducting   tunnel  junctions  operating   at   low
temperatures\cite{ref18}. For high frequencies the mechanism of the
THz-photons detection in superlattices was described\cite{ref35} as
a   bulk   superlattice   effect  caused   by   dynamical
localisation of electrons.

\begin{figure}
\epsfxsize=120mm
\epsffile{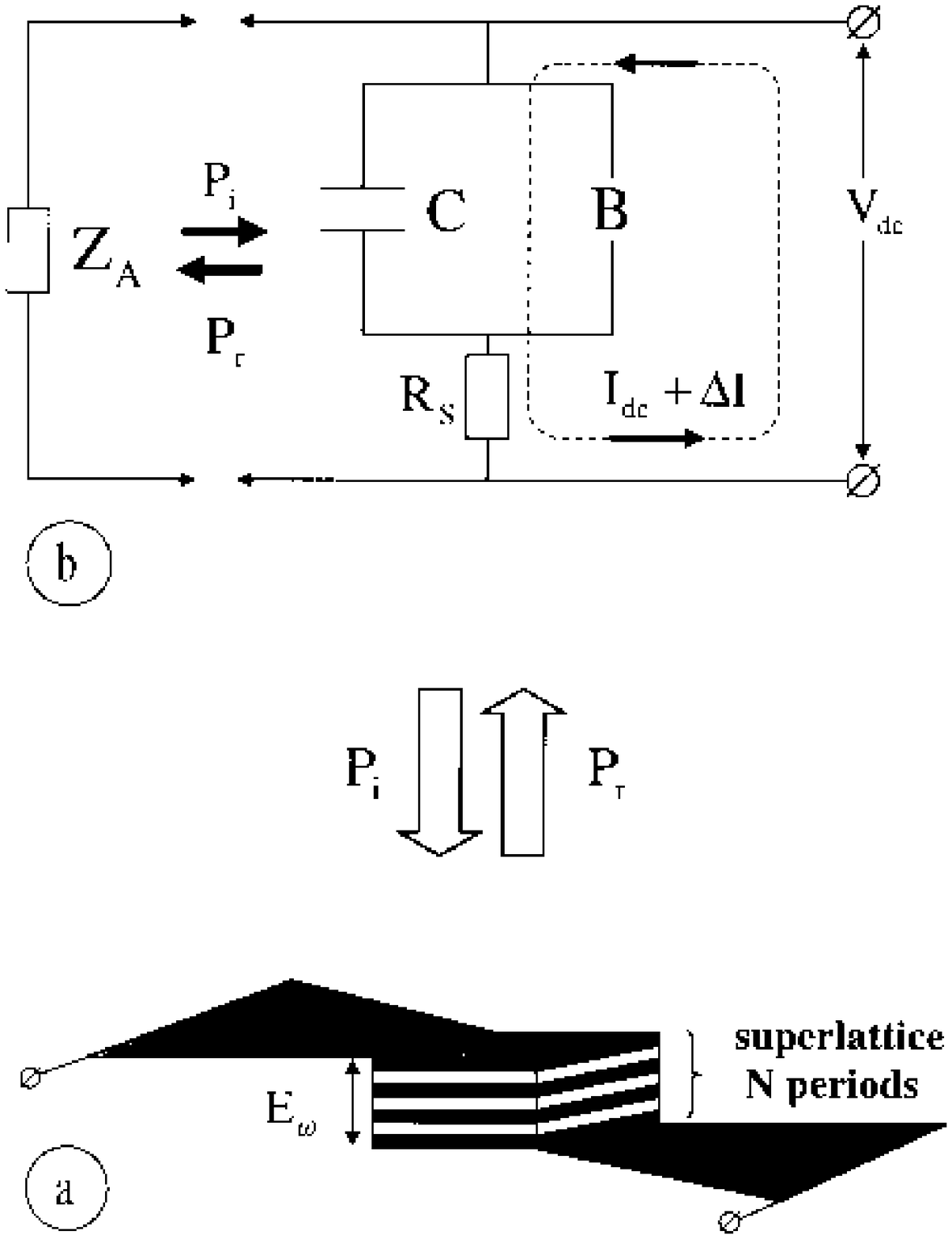}
\caption{    
(a)  THz-radiation  coupled  to  a  $N$-period
semiconductor superlattice by a co-planar broad band bow-
tie  antenna,  $P_i$ and $P_r$  are the incident and reflected powers
respectively.  (b) Equivalent circuit  for  a  THz-photon
detector  with  a  dc  voltage  bias  source:  $B$--miniband
electrons  capable  to  perform  Bloch  oscillations,  
$C$--superlattice capacitance, $R_S$--parasitic series  resistance,
$Z_A$--bow-tie antenna impedance, $V_{dc}$--dc bias voltage.
}
\label{fig1}
\end{figure}

    In  the  present  paper we develop a self-consistent
theory of the superlattice current responsivity. We apply
the   Boltzmann  equation  approach  for  describing  the
electron motion in the superlattice miniband\cite{ref3} and  assume
an  equivalent circuit for the superlattice coupled to  a
broad-band antenna (see Fig.\ \ref{fig1}), which is similar to  the
equivalent  circuit  used  in  resonant-tunnelling\cite{ref36}  and
Schottky  diodes\cite{ref37} simulations. The suggested  equivalent
circuit of the device allows one to treat microscopically
the  high-frequency  response of the  miniband  electrons
and,  simultaneously, take into account a finite matching
efficiency   between  the  detector   antenna   and   the
superlattice  in  the presence of parasitic  losses.  Our
analytic  results  lead  to  the  identification  of   an
important  physical  concept: the  excitation  of  hybrid
plasma-Bloch\cite{ref38}  oscillations 
{\it in the  region  of  positive
differential   conductance  of  the   superlattice}.   The
numerical  computations, performed for  room  temperature
behaviour  of  currently available  superlattice  diodes,
show   that   both  the  magnitudes  and   the   roll-off
frequencies  of the responsivity are strongly  influenced
by  this  effect.  The  excitation  of  the  plasma-Bloch
oscillations gives rise to a resonant-like dependence  of
the  responsivity  on  the incident radiation  frequency,
improving essentially the coupling of the superlattice to
the detector antenna. We will also show that peak current
densities  in  the device and its geometrical  dimensions
should  be  properly optimised in order  to  get  maximum
responsivity for each frequency of the incident  photons.
Finally,  we  will  present numerical  estimates  of  the
responsivity for the 1-4 THz frequency band  and  compare
its  value  with  the  quantum efficiency  $e/\hbar\omega$ 
of  an  ideal
detector.

\section{DESCRIPTION OF THE MODEL}

    For the description of the ac electron transport in a
superlattice  we  use  a  quasi-  classical  wave  packet
treatment  of  the  electron motion in a 
superlattice\cite{ref1,ref2,ref3}.
The  energy spectrum of electrons in a miniband is  taken
in a tight-binding approximation:
\begin{equation}
\epsilon({\bf p}) = {\Delta\over 2}
\left[ 1 - \cos\left({p_z d\over\hbar}\right)\right]
+{p_x^2+p_y^2\over 2 m}
\label{(1)}
\end{equation}
where $\Delta$  is  the  superlattice miniband width,  $d$  is  the
superlattice  period,  $p_z$ is  the  quasi-momentum   of   an
electron  along  the superlattice axis (perpendicular  to
the layers),  $p_x,p_y$ are the quasi-momentum components along the
superlattice  layers,  and  $m$ is  the  effective  mass  of
electrons along the superlattice layers.

    The  quasi-classical velocity $v_z(p_z)$ of an electron moving
along  the  superlattice axis and the time-derivative  of
the quasi-momentum are given by the expressions
\begin{eqnarray}
v_z(p_z) &=& {\partial\epsilon({\bf p})\over\partial p_z}
= v_0 \sin \left({p_z d\over\hbar }\right)\\
{\dot p}_z &=& e E_z(t)
\end{eqnarray}
where $e$  is  the electron charge,  $v_0 = \Delta d/2\hbar$
is the maximum velocity
of  electrons along the superlattice axis,  and $E_z(t)$  is  the
time-periodic   electric   field   directed   along   the
superlattice axis.

     The electric current density $j_z(t)$ is calculated from the
distribution function,
\begin{equation}
j_z(t) = e \int v_z(p_z) f({\bf p},t) {2d{\bf p}\over(2\pi\hbar)^3}\;,
\label{(4)}                                                        
\end{equation}
which satisfies the time-dependent Boltzmann equation
\begin{equation}
{\partial f({\bf p},t)\over\partial t}
+ e E_z(t)
{\partial f({\bf p},t)\over\partial p_z} =
\left( {\partial f\over\partial t} \right)_{\rm coll}
\label{(5)}
\end{equation}

     Below  we use the relaxation-time approximation  for
the collision integral\cite{ref1,ref2,ref3}
\begin{equation}
\left( {\partial f\over\partial t} \right)_{\rm coll}
= - {f({\bf p},t) - f_0(p)\over \tau}
\label{(6)}
\end{equation}
where   $\tau$ is  the  constant relaxation time for  electron's
scattering,   and   $f_0(p)$ is   the  equilibrium   distribution
function.

     The relaxation-time approximation is, of course,  an
oversimplification  of the numerous scattering  processes
taking   place  in  a  real  superlattice.  Nevertheless,
several papers\cite{ref39} have demonstrated that the phenomenon of
dynamical localisation can be described very well  within
this  approximation  when compared to  the  corresponding
results    obtained   from   a   full-scale   Monte-Carlo
simulation.  This  circumstance  lends  support  to   the
simplified  model for the collision integral.  The  great
advantage  of  Eqs.(\ref{(5)}) and (\ref{(6)}) is that  they  allow  an
analytical calculation of the time-dependent current,  to
be  used in the equivalent circuit analysis, and thus  we
can  study  in detail various parametric dependencies  of
the calculated quantities.

     We  would also like to emphasise here that the  wave
packet  description of electron motion in a  superlattice
is valid if the following inequalities are 
fulfilled\cite{ref3,ref40}:
\begin{eqnarray}
eEd &\ll& \Delta,\Delta_G\\
\hbar\omega&\ll& \Delta,\Delta_G\\
\hbar\nu&\ll& \Delta,\Delta_G
\end{eqnarray}
were $\Delta_G$  is the width of the superlattice minigap, 
$\nu = 1/\tau$ is  the
electron's    collision   frequency,   and  $\omega$    is    the
characteristic  frequency of the external  ac  field.  We
also  assume  that  the electronic  mean  free  path 
$\ell_{\rm FP} = v_0\tau = \Delta d / 2 \hbar \nu$  is
smaller than the superlattice length  $L$
in order to neglect
the influence of the boundaries on the superlattice high-
frequency properties.

\section{THEORETICAL FORMALISM}

\subsection{Path integral}
The exact solution of Eqs. (\ref{(5)}) and (\ref{(6)}) for arbitrary
time-dependent  electric field  can be presented  in  the
form of a path integral\cite{ref41}:
\begin{equation}
f({\bf p},t) = \int_{-\infty}^t \nu dt_1
\exp\left[ - \nu (t - t_1)\right ]
f_0(p_z - \int_{t_1}^t eE_z(t_2) dt_2)
\label{(10)}
\end{equation}
Using  Eqs.(\ref{(4)})  and  (\ref{(10)})  we find  the  time-dependent
current  $I(t)$ describing ac transport in a superlattice,  with
electron  performing  ballistic  motion  in  a  mini-band
according  to  the  acceleration  theorem  and  suffering
scattering\cite{ref2,ref3}:
\begin{equation}
I(t) = 2I_p \int_{-\infty}^t \nu dt_1
\exp\left[ -\nu (t-t_1)\right]
\sin\left[ {e\over N\hbar} \int_{t_1}^t V(t_2)dt_2\right ]
\label{(11)}
\end{equation}
where $V(t) = LE_z(t)$ is   the   voltage   across   the   superlattice
perpendicular to the layers,  $L = N d$ is the superlattice length,
$N$ is  the  number of periods in the superlattice sample, 
$I_p = S j_p$, $S = \pi a^2$
is  the  area  of the superlattice,  $a$ is the  superlattice
mesa radius, and
\begin{equation}
j_p = e {v_0\over 2} \int \cos \left({p_z d\over \hbar}\right)
f_0(p) {2d{\bf p}\over (2\pi\hbar)^3}
\label{(12)}
\end{equation}
is  the  characteristic current density. The  integration
over $p_z$  in Eq.(\ref{(12)}) must be carried out over the Brillouin
zone  $-\pi\hbar/d\le p_z \le \pi\hbar/d$.

     In  particular,  for a dc voltage  $V_{SL}$ applied  to  the
superlattice (Eq.\ref{(11)}) results in the Esaki-Tsu\cite{ref1}  current-
voltage curve
\begin{equation}
I^{SL}_{DC}(V_{SL}) = 2I_P 
{(V_{SL}/V_P)\over
1+(V_{SL}/V_P)^2}
\label{(13)}
\end{equation}
At peak voltage 
$V_{SL} = V_P =N\hbar\nu/e$ 
or, equivalently, at peak electric field
$E_{SL}=E_P=\hbar\nu/ed$,  
the  current in the superlattice reaches  its  maximum
(peak)  value  $I_P$ ,  so that $j_P$ can be defined  as  the  peak
current density.

      The   peak  current  density  $j_P$ and  the  scattering
frequency  $\nu$ can  be considered as the main parameters  of
the  employed  model. They can readily be estimated  from
experimentally  measured or numerically simulated  values
of $I_P$  and $V_P$. For both degenerate and non-degenerate electron
gas one gets \cite{ref2,ref3}
\begin{equation}
j_P = en {v_0\over 2}
\label{(14)}
\end{equation}
if  $\Delta\gg kT,\epsilon_F$,
where $kT$ is the equilibrium thermal excitation energy,
$\epsilon_F = \hbar^2(3 \pi^2 n)^{2/3}/(2m_{\rm eff}) $
is  the  Fermi  energy of degenerate electrons, 
$n=\int f_0({\bf p}) 2d{\bf p}/(2\pi\hbar)^3$ is  the
charge   carrier  density,   
$m_{\rm eff} = m_{zz}^{1/3} m^{2/3}$
is  the  density  of  states
effective  mass  near the miniband bottom,  and 
$m_{zz} = 2\hbar^2 /\Delta d^2$  is  the
effective mass of electrons along the superlattice  axis.
In  the  particular  case  of the  Boltzmann  equilibrium
distribution function Eq.(\ref{(12)}) yields\cite{ref3}
$j_P=(env_0/2) [I_1(\Delta/2kT)/I_0(\Delta/2kT)]$, 
where $I_{0,1}$  are  the
modified Bessel functions.

\subsection{Monochromatic excitation}

     We now suppose that in addition to the dc voltage  $V_{SL}$,
an   alternating  sinusoidal  voltage  with   a   complex
amplitude $V_\omega$ is applied to the superlattice:
\begin{equation}
V(t) = V_{SL} + {1\over 2}
\left[V_\omega \exp(i\omega t) +
V_\omega^* \exp(-i\omega t)\right]
\label{(15)}
\end{equation}
Generally,  $V_{SL}, V_\omega$
can  be  found  from  an  analysis  of   the
equivalent circuit given in
Fig.\ \ref{fig1}. We write the ac voltage amplitude as 
$V_\omega = |V_\omega| e^{i\psi}$; 
both $|V_\omega|$   and $\psi$
can  be  obtained  self-consistently  taking  account  of
reflection  of the THz photons from the superlattice  and
their absorption in the series resistor $R_S$.

    Making use Eq.(\ref{(11)}) we obtain\cite{ref3}:
\begin{equation}
I(t) = 2I_P \int_0^{\infty} \nu dt_1 \exp(-\nu t_1)
\sin\left[ {eV_{SL}\over N\hbar} t_1 + \Phi(t,t_1) \right ]
\label{(16)}
\end{equation}
where
\begin{equation}
\Phi(t,t_1) = {e\over N\hbar\omega} \times
{1\over 2} \left\{ iV_\omega\exp(i\omega t)
[\exp(-i\omega t_1)-1] + c.c. \right \}
\label{(17)}
\end{equation}
According  to  Eq.(\ref{(16)}),  electrons  in  a  superlattice
miniband  perform  damped  Bloch  oscillations  with  the
frequency
$\Omega_B = eV_{SL} / N\hbar = eE_{SL} d/\hbar$, 
and the phase $\Phi(t,t_1)$ modulated by the external  ac
voltage.

     In the limit of  $V_\omega \to 0$
Eq.(\ref{(16)}) reduces to the dc 
current-voltage curve given by Eq.(\ref{(13)}). On the other hand,  when
the frequency of the ac voltage is small, $\omega\tau\ll 1$, 
we get
\begin{equation}
\Phi(t,t_1) = {et_1\over N\hbar} \times
{1\over 2} \left [ V_\omega \exp(i\omega t) + c.c.\right]
\label{(18)}
\end{equation}
which  corresponds  to  a slow modulation  of  the  Bloch
frequency  by external voltage. In this case the  current
in  the  superlattice follows instantaneously  the  time-
dependent  ac voltage according to the dc current-voltage
curve.

      Equation  (\ref{(16)})  contains,  as  special  cases   the
following  results:  (i) a harmonic   voltage   $V(t)$
($V_{SL}=0$) leads  to
dynamical  localisation, and current harmonics generation
with  oscillating power dependence\cite{ref3}; (ii) a  dc  current-
voltage  characteristics  of the irradiated  superlattice
$I_{DC}(V_{SL},V_\omega) = (\omega/2\pi) \int I(t) dt$
shows  resonance  features (`Shapiro steps')  leading  to
absolute   negative  conductance\cite{ref3,ref5,ref17};  
(iii)   and   to
generation  of dc voltages (per one superlattice  period)
that are multiples of $\hbar\omega/e$\cite{ref13}.

\subsection{Method of perturbations}

     Let  us assume that the external ac voltage $V_\omega$   is  so
small  that  perturbation  theory  holds,  while  the  dc
voltage  $V_{SL}$ applied  to the superlattice keeps  its  finite
value. Expanding Eq. (\ref{(16)}) around $V_{SL}$  at
$V_\omega\to 0$  in a Taylor series,
we  obtain  the  time-dependent electric current  in  the
form:
\begin{equation}
I(t) = I_{DC}^{SL}(V_{SL})
+ {1\over 2}
\left[ G^{SL}_{AC}(\omega,V_{SL}) V_\omega e^{i\omega t} + c.c.\right]
+\Delta I^{SL}_{DC}(\omega,V_{SL})
\label{(19)}
\end{equation}
where
\begin{equation}
G^{SL}_{AC}(\omega,V_{SL}) = G_0 F_1(\omega,V_{SL})
\label{(20)}
\end{equation}
is the superlattice ac conductance\cite{ref3},  
$G_0 = 2I_P/V_P$ is the superlattice
conductance at $\omega\to 0, V_{SL}\to 0$,
\begin{equation}
F_1(\omega,V_{SL}) = 
{1 + i\omega\tau - (V_{SL}/V_P)^2 
\over
\left [1+(V_{SL}/V_P)^2 \right]
\left[(1+i\omega\tau)^2 + (V_{SL}/V_P)^2\right] }
\label{(21)}
\end{equation}
is  a dimensionless function describing the dependence of
the   superlattice  conductance  both  on  frequency  and
applied dc voltage (field), and
\begin{equation}
\Delta I^{SL}_{DV}(\omega,V_{SL}) = 
{1\over 4} |V_\omega|^2 F_2(\omega,V_{SL})
\label{(22)}
\end{equation}
where
\begin{eqnarray}
F_2(\omega,V_{SL}) &=&
{I_{DC}^{SL}(V_{SL}+N\hbar\omega/e)
- 2 I_{DC}^{SL}(V_{SL})+
I_{DC}^{SL}(V_{SL}-N\hbar\omega/e)
\over
(N\hbar\omega/e)^2}\nonumber\\
&=&
-{4I_P\over V_P^2}
{(V_{SL}/V_P)
\left[3 + (\omega\tau)^2 - (V_{SL}/V_P)^2\right]
\over
[1+(V_{SL}/V_P)^2]
[1+(V_{SL}/V_P+\omega\tau)^2]
[1+(V_{SL}/V_P-\omega\tau)^2]}
\label{(23)}
\end{eqnarray}
is  the change in the dc current in a superlattice caused
by THz-photons\cite{ref3}.

   At $\omega\to 0$ Eq.(\ref{(20)}) yields the dc differential conductance of
the superlattice
\begin{eqnarray}
G_{AC}^{SL}(\omega\to 0,V_{SL}) &=& dI_{DC}^{SL}(V_{SL})/ dV_{SL}\nonumber\\
&=& G_0 
{1-(V_{SL}/V_P)^2 \over
\left[ 1 + (V_{SL}/V_P)^2 \right]^2}
\label{(24)}
\end{eqnarray}
while  at $V_{SL}\to 0$
one gets the well known Drude formula for  the
ac conductivity of the electron gas $\sigma(\omega)$
\begin{equation}
\sigma(\omega) = {\sigma_0\over 1 + i\omega\tau}
\label{(25)}
\end{equation}
where the small-field dc conductivity of the superlattice $\sigma_0$
can  be  given  in  terms  of the  dc  conductance  
$\sigma_0 = G_0 L / S$  or,
equivalently,  in terms of peak current density  and  the
peak electric field
\begin{equation}
\sigma_0 = 2 {j_P \over E_P}
\label{(26)}
\end{equation}
    According  to  Eqs.(\ref{(15)})  and  (\ref{(19)})  the  ac  power
 $P^{SL}_{\rm abs}$ absorbed in the superlattice is
\begin{equation}
P^{SL}_{\rm abs} = {1\over 2}
{\rm Re} G^{SL}_{AC}(\omega,V_{SL}) |V_\omega|^2
\label{(27)}
\end{equation}
where  the real part of the superlattice conductance  can
be  presented  in  analogy with  Eq.  (23)  in  a  finite
difference form\cite{ref3,ref42}
\begin{eqnarray}
{\rm Re}G^{SL}_{AC}(\omega,V_{SL}) &=& {e\over 2N\hbar\omega}
\left[I^{SL}_{DC}(V_{SL} + N\hbar\omega/e)
-I^{SL}_{DC}(V_{SL} - N\hbar\omega/e)\right]\nonumber\\
&=& {2I_P\over V_P}
{1+(\omega\tau)^2 - (V_{SL}/V_P)^2 \over
[1+(V_{SL}/V_P + \omega\tau)^2]
[1+(V_{SL}/V_P - \omega\tau)^2]}
\label{(28)}
\end{eqnarray}
    We  note here that our calculations assume a uniform
dc/ac  electric  fields  inside  the  superlattice.  This
assumption is justified for the sub-threshold dc electric
fields $V_{SL}\le V_P$, 
and/or for ac field frequencies larger than  the
transit-time  frequency  of space-charge  waves  
$\omega\ge v_0/L$ in  the
superlattice\cite{ref38}.

\subsection{Stimulated emission and positive photo-current}

      Equation  (\ref{(28)})  for  the  superlattice  conductance
demonstrates a possibility of the resonant interaction of
the  THz-photons with electrons performing  damped  Bloch
oscillations in the superlattice. For  $\omega\tau\gg 1$
the condition  for
the  resonance  has  a  form 
$\omega = \pm (V_{SL}/V_P)\tau$ which can  be  equivalently
expressed  as $\omega=\pm\Omega_B$. 
At dc electric fields 
$E_{SL}\ge E_P\sqrt{1+(\omega\tau)^2}$ 
the real  part  of
the  superlattice conductance starts to be negative which
corresponds to negative absorption (stimulated  emission)
of  the THz-photons. In the limit $\omega\tau\gg 1$ 
photons with frequency $\omega>\Omega_B$
are absorbed and the ones having frequency
$\omega < \Omega_B$  are emitted.

     Equation (\ref{(22)}) describes the dc current change in the
superlattice under the influence of the THz-photons.  The
standard result of the classical rectification theory\cite{ref18}
\begin{equation}
\Delta I^{SL}_{DC}(\omega\to 0,V_{SL})=
{1\over 4} |V_\omega|^2
{I_{DC}^{SL}}^{''}(V_{SL})
\label{(29)}
\end{equation}
where
\begin{equation}
{I^{SL}_{DC}}^{''}(V_{SL})
= -4 {I_P \over V_P^2}
{(V_{SL}/V_P)\left[3 - (V_{SL}/V_P)^2\right] \over
\left[ 1 + (V_{SL}/V_P)^2 \right]^3 }                       
\label{(30)}
\end{equation}
is the second derivative of the dc current-voltage curve
of  the superlattice, is obtained from Eq. (\ref{(22)}) when  the
electron system relaxes during one period of the ac field.

     At  dc  bias  fields  
$E_{SL} < E_P \sqrt{3+(\omega\tau)^2}$ 
the  current  change  in  the
irradiated superlattice is negative which corresponds  to
the  onset of suppression of the current due to dynamical
localisation  of  carriers
\cite{ref3,ref4,ref5,ref6,ref7,ref8,ref9,ref10,ref11,ref12,ref13,ref14,ref15}. 
In the opposite  case
$E_{SL} > E_P \sqrt{3+(\omega\tau)^2}$   
a
positive  current change (positive photocurrent)  occurs.
It  is important to note that the condition of occurrence
of  the positive photo-current implies, according to 
Eqs.(\ref{(22)})  and  (\ref{(27)}), the existence of stimulated emission  of
photons from the superlattice, $P^{SL}_{\rm abs}<0$.

     The  increase  of  the dc current  in  superlattices
caused by stimulated emission of THz-photons has recently
been   observed  in  experiments\cite{ref33}.  These   experimental
results  lend support to the validity of the  theoretical
formalism employed in the present paper for the  analysis
of the superlattice THz-field response.

\section{CURRENT RESPONSIVITY}
\subsection{Equivalent circuit calculation}

      We   define  the  current  responsivity\cite{ref18}  of   the
superlattice detector as the current change $\Delta I$  induced  in
the external dc circuit per incoming ac signal power $P_i$.
\begin{equation}
R_i(\omega,V_{SL})= {\Delta I \over P_i}
\label{(31)}
\end{equation}
This  definition  takes into account both  the  parasitic
losses  in  the  detector and the finite  efficiency  for
impedance-matching  of  the  incoming  signal  into   the
superlattice  diode.  In  the  further  analysis  of  the
current  responsivity we use both the frequency $f = \omega/2\pi$ and  the
angular frequency $\omega$ notations.

     According to Eqs.\ref{(22)} and \ref{(27)}, in the small-signal
approximation both the dc current change 
$\Delta I^{SL}_{DC}$ and  the  power $P^{SL}_{\rm abs}$
absorbed  in  the  superlattice are proportional  to  the
square modulus of the complex voltage $|V_\omega|^2$. 
This circumstance
permits  us  to  calculate self-consistently  $|V_\omega|^2$ for  given
values  of  the incoming power  making use  a  linear  ac
equivalent  circuit analysis and, then, find the  current
responsivity $R_i(\omega,V_{SL})$.

     The  results  of the calculation of the superlattice
current  responsivity   $R_i(\omega,V_{SL})$
are presented  in  the  following
form:
\begin{equation}
R_i(\omega,V_{SL})=
{R_i^{(0)}(\omega,V_{SL}) A(\omega,V_{SL}) \over
1+ R_S (dI^{SL}_{DC}(V_{SL})/dV_{SL}) }
\label{(32)}
\end{equation}
where
\begin{equation}
R_i^{(0)}(\omega,V_{SL}) = -{e\over N \hbar\nu}
{(V_{SL}/V_P)[3 + (\omega\tau)^2 - (V_{SL}/V_P)^2]\over
[1+(V_{SL}/V_P)^2]
[1+(\omega\tau)^2-(V_{SL}/V_P)^2]}
\label{(33)}
\end{equation}
is the superlattice current responsivity under conditions
of  a  perfect  matching and neglecting parasitic  losses
($R_S\to 0$)\cite{ref35} ,where $R_S$ is the series resistance.

     The factor $A(\omega,V_{SL})$  
in Eq. (\ref{(32)}) describes the effect of  the
electrodynamical  mismatch between the  antenna  and  the
superlattice  and  the signal absorption  in  the  series
resistance
\begin{equation}
A(\omega,V_{SL})=
\left[1 - 
\left|{
Z_A-(Z^{SL}_{AC}(\omega,V_{SL}) + R_S) \over
Z_A+(Z^{SL}_{AC}(\omega,V_{SL}) + R_S)}
\right|^2
\right]\times
{{\rm Re}Z^{SL}_{AC}(\omega,V_{SL})\over
{\rm Re}Z^{SL}_{AC}(\omega,V_{SL})+R_S}\;.
\label{(34)}
\end{equation}
The first factor in Eq. (\ref{(34)}) describes the reflection  of
the  THz-photons due to mismatch of the antenna impedance $Z_A$
and  the total impedance of the device
$Z^{SL}_{AC}(\omega,V_{SL})+R_S$, with the second
one  being responsible for sharing of the absorbed  power
between  the active part of the device described  by  the
impedance $Z^{SL}_{AC}(\omega,V_{SL})$ and the series resistance $R_S$.

    The superlattice impedance is defined as
\begin{equation}
Z^{SL}_{AC}(\omega,V_{SL})=
1/\left[G^{SL}_{AC}(\omega,V_{SL}) + i\omega C\right]
\label{(35)}
\end{equation}
where $G^{SL}_{AC}(\omega,V_{SL})$ is the superlattice conductance,
$C=\epsilon_0 S / 4\pi L$  is the capacitance
of  the  superlattice,  and  
$\epsilon_0$ is  the  average  dielectric
lattice constant.

     Finally,  the last factor in Eq. (\ref{(32)}) describes  the
redistribution of the external bias voltage  $V_{DC}$
between  the
dc  differential resistance of the superlattice  
$(dI^{SL}_{DC}(V_{SL})/dV_{SL})^{-1}$ and  the
series  resistance  $R_S$, with the dc voltage  drop  on  the
superlattice  $V_{SL}$ being determined by the  solution  of  the
well-known load equation\cite{ref18}
\begin{equation}
V_{DC}=V_{SL} + I^{SL}_{DC}(V_{SL})R_S\;.
\label{(36)}
\end{equation}

\subsection{Classical rectification at high frequencies}

    Suppose now that the following set of inequalities is
satisfied
\begin{eqnarray}
\omega\tau &\ll& 1\\
R_S&\ll& R_0\\
\omega&\gg& 1/\sqrt{R_0 R_S} C\;,
\end{eqnarray}
which   implies  that  the  frequency  of  the   incident
radiation  $\omega$ is  small in comparison with  the  scattering
frequency of electrons $\nu$, series resistance $R_S$ is small  in
comparison  with  the small-field dc  resistance  of  the
superlattice $R_0=1/G_0$ and the appropriate $RC$ time of the  device
is high enough. In this case Eq. (\ref{(32)}) gives
\begin{equation}
R_i(\omega,V_{SL}) = 
{2I^{SL}_{DC}}^{''}(V_{SL})\times
{Z_A\over (1+ Z_A/R_S)^2}
\times
{1\over (\omega R_S C)^2}
\label{(40)}
\end{equation}
Equation (\ref{(40)}) was obtained by Sollner et.al.\cite{ref36} following
Torrey's  and Whitmer's approach\cite{ref37} developed for analyses
of  classical crystal rectifiers. They also took  account
of  the  mismatch between the antenna impedance $Z_A$ and  the
device impedance. It was applied to analyse detection  of
photons  by  resonant  tunnelling  diodes  in  the   THz-
frequency  range. This expression demonstrates  that  the
responsivity   of   the  classical  rectifier   at   high
frequencies  is proportional to the second derivative  of
the   dc   current-voltage  curve,  and  hence   strongly
decreases with increasing of frequency, 
$R_i(\omega,V_{SL}) \propto \omega^{-2}$.

      The  factor 
      $1/R_S C$  in  Eq.  (\ref{(40)})  defines  the  roll-off
frequency for the current responsivity and, consequently,
capacitance of the device should be minimised in order to
increase  the  responsivity in the high frequency  range.
The fast decrease of the responsivity with increasing  of
frequency can be attributed to imperfect matching of  the
device  to  the  antenna impedance when incoming  photons
either  are reflected from the device or are absorbed  in
the series resistor.

      In  our  case  Eq.  (\ref{(40)})  shows  that  the  current
responsivity  of the superlattice can reach  its  maximum
(negative) value at bias voltage $V_{SL}\simeq 0.4 V_P$, 
i.e. in the region  of
the  positive  conductance. Further, the responsivity  of
the  superlattice does not depend on the  length  of  the
device because  $V_P\propto L$ and 
$C\propto 1/L$, and, finally, the responsivity  is
proportional  to  the current density $j_P$. Consequently,  in
order  to increase the responsivity in the high-frequency
range  one  should employ highly conductive  superlattice
samples  for  which  inequalities (37)-(39)  may  be  not
satisfied.   In   this  case  the  interaction   of   the
superlattice   with  the  incoming   radiation   can   be
essentially  changed by excitation of the  eigenmodes  in
the superlattice device.

\subsection{Superlattice  dielectric function.  Hybridisation  of
Bloch and plasma oscillations}

    Let us analyse the condition of optimised matching of
the superlattice to the incident radiation going back  to
the  general  Eq. (\ref{(32)}). Assuming the limit of  negligible
series  resistance $R_S\to 0$ this condition can be  obtained  from
solution of the equation
\begin{equation}
Z^{SL}_{AC}(\omega,V_{SL})=Z_A
\label{(41)}
\end{equation}
for  the complex frequency
$\omega(V_{SL})$ . This solution determines the
resonant  line position and the line width at  which  the
absorption  in  the  superlattice tends  to  its  maximum
value.

     Using Eqs. (\ref{(20)}) and (\ref{(35)}) one can transform 
Eq. (\ref{(41)})
to the following form:
\begin{equation}
\epsilon(\omega,E_{SL}) = 
{\epsilon_0\over i\omega C Z_A}
\label{(42)}
\end{equation}
where
\begin{equation}
\epsilon(\omega,E_{SL}) = \epsilon_0
+ {4\pi \sigma_0\over i \omega}
F_1(\omega,E_{SL})
\label{(43)}
\end{equation}
is  the dielectric function of the superlattice, with the
dc  field  $E_{SL}$ being applied to the device\cite{ref38},  
$F_1(\omega,E_{SL})$ is defined  by
Eq. (\ref{(21)}).

    In the high-frequency limit  
$\epsilon_0/CZ_A\omega \to 0$ the solution of Eq. (\ref{(41)})
coincides with the solution of the equation
\begin{equation}
\epsilon(\omega,E_{SL})=0
\label{(44)}
\end{equation}
describing  eigenfrequencies  $\omega^H_\pm$
of the hybrid  plasma-Bloch
oscillations in a superlattice\cite{ref38}
\begin{equation}
\omega^H_\pm(E_{SL}) = \pm \omega_P
\left [ {1\over 1 + (E_{SL}/E_P)^2}
+ \left( {\nu\over\omega_P}\right)^2
(E_{SL}/E_P)^2 \right]^{1/2} + i\nu
\label{(45)}
\end{equation}
where  $\omega_P$  is  the  plasma  frequency  of  electrons  in   a
superlattice. The plasma frequency $\omega_P$ 
can be given in  terms
of  the small-field dc conductivity $\sigma_0$ or, equivalently, in
terms of the peak current density $j_P$
\begin{equation}
\omega_P =
\left({4\pi\sigma_0 \nu\over\epsilon_0}\right)^{1/2} =
\left({8\pi j_P ed\over\epsilon_0\hbar}\right)^{1/2}
\label{(46)}
\end{equation}
Equation  (\ref{(46)})  reduces in the particular case  of 
wide-miniband superlattices ($\Delta\gg kT,\epsilon_F$) 
to the standard formula 
$\omega_P = (4\pi e^2 n / \epsilon_0 m_{zz})^{1/2}$.

     In  the  limiting case of small applied dc  electric
fields  $E_{SL}/E_P \to 0$ one  finds from 
Eq. (\ref{(45)}) the plasma frequency
$\omega^H_\pm \to \pm\omega_P$,
while  in  the  opposite case 
$E_{SL}/E_P \to \infty$, the Bloch  frequency 
$\omega^H_\pm\to\pm\Omega_B=\pm eE_{SL} d/\hbar$  is
recovered.  The  scattering frequency  $\nu$ in  Eq.  (\ref{(45)})  is
responsible  for  the  line  width  of  the  plasma-Bloch
resonance.

\begin{figure}
\epsfxsize=120mm
\epsffile{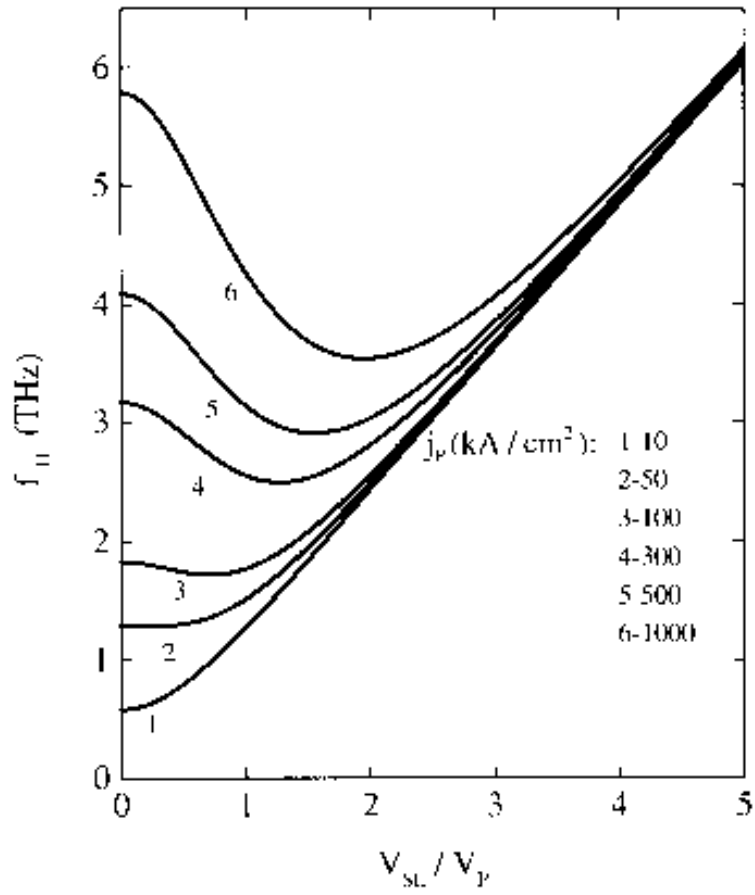}
\caption{
The calculated hybrid plasma-Bloch oscillation
frequency  $f_H$ is  plotted as a function of  the  normalised
superlattice  voltage drop $V_{SL}/V_P$ for different values  of  the
peak  current densities $j_P=$ 10, 50, 100, 300, 500,  and  1000
kA/cm$^2$. Typical values of the superlattice parameters ( $d$
= 50{\AA},  $E_P$=10 kV/cm, $\epsilon_0$ =13) were used for the calculations.
}
\label{fig2}
\end{figure}

      We   have   calculated   the  hybrid   plasma-Bloch
oscillation frequency $f_H=\omega^H_+/2\pi$, 
using Eqs. (\ref{(45)}) and (\ref{(46)}), 
for the
typical  values of the superlattice 
parameters\cite{ref27,ref28,ref29}  
$\epsilon_0\simeq 13$, 
$d\simeq 50${\AA},
$E_P\simeq 10$ kV/cm,
$f_\nu = \nu/2\pi = 1.2$THz for
different values of the current densities  $j_p$ (Fig.\ \ref{fig2}).  For
small  values of the current densities $j_P \simeq 10$ kA/cm$^2$
the frequency  of
the hybrid oscillation increases with applied voltage  in
all  range  of  the parameter
$V_{SL}/V_P$ . On the  other  hand,  for
higher  values  of  the  current  densities
$j_P\simeq (50 - 1000)$ kA/cm$^2$   the  hybrid
oscillation's   frequency   starts   to   decrease   with
increasing  bias  voltage  in the  sub-threshold  voltage
range  $V_{SL}\le V_P$.  Then, at super-threshold voltages
$V_{SL}\ge V_P$,  
$\omega_H$ starts  to
increase  again  tending to the Bloch  frequency.  It  is
important  to note that at high values of the dc  current
densities  $j_P$ the  hybrid plasma-Bloch oscillations  become
well  defined eigenmodes of the system ($f_H \ge f_\nu$). 
Therefore,  an
essential improvement of the matching efficiency  between
antenna and the superlattice can be expected in the high-
frequency  range  due to a resonant  excitation  of  this
eigenmode in the device.

\subsection{High-frequency limit}

    In the high-frequency case, when the signal frequency
$f=\omega/2\pi$
is  larger than the scattering frequency
$f_\nu = \nu/2\pi$, description  of
the   superlattice  response  based  on   the   classical
rectification  theory is no longer valid. Let  us  employ
here  Eq.  (\ref{(33)})  in  order to analyse the  high-frequency
limit  of  the  responsivity in the  ideal  case  of  the
perfect matching and neglecting the series resistance. At
$\omega\gg\nu,\Omega_B$, 
i.e. in the frequency band were ac field absorption and
negative   photo-current   are   predominated   in    the
superlattice  response, Eq. (\ref{(33)})  yields  the  frequency-
independent value for the current responsivity
\begin{equation}
R^{(0)}_i(\omega\to\infty,V_{SL}) =
-{e\over N\hbar\nu} 
{V_{SL}/V_P\over
1+(V_{SL}/V_P)^2}
\label{(47)}
\end{equation}
At positive bias
$V_{SL}/V_P>0$  the responsivity is negative (reduction
of the dc current occurs due to dynamical localisation of
carriers
\cite{ref3,ref4,ref5,ref6,ref7,ref8,ref9,ref10,ref11,ref12,ref13,ref14,ref15}).   
The  dc  voltage   dependence   of   the
responsivity  reproduces the dc Esaki-Tsu current-voltage
curve\cite{ref1}. The responsivity reaches its maximum value 
$R^{(0)}_{i{\rm max}} =
-e/(2N\hbar\nu) = -1/(2 V_P)$ at $V_{SL}=V_P$.
For high values of the applied voltages 
$V_{SL}\gg V_P$ we get 
$R^{(0)}_i(\omega\to\infty,V_{SL}\to\infty)=-1/V_{SL}$.

     Let  us  compare  the high-frequency  limit  of  the
responsivity  of  the  superlattice  with   the   quantum
efficiency $R_{\rm max} = e/\hbar\omega$    
which  is  believed  to  be  a  fundamental
restriction for the responsivity of superconductor tunnel
junctions\cite{ref18}.  This quantum efficiency (or quantum  limit)
corresponds to the tunnelling of one electron across  the
junction  for  each  signal  photon  absorbed\cite{ref18},  with  a
positive sign of the responsivity.

\begin{figure}
\epsfxsize=120mm
\epsffile{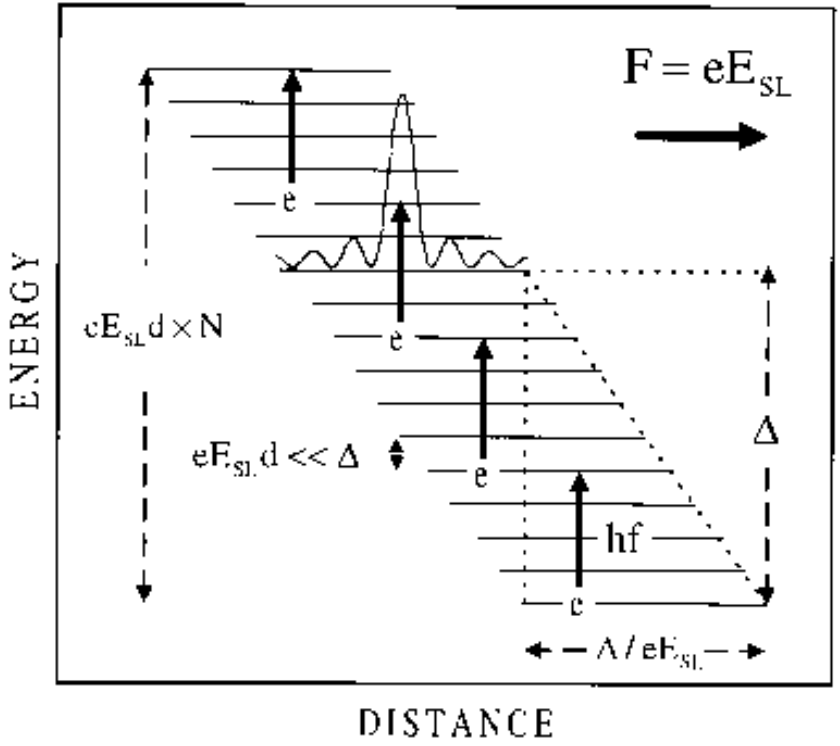}
\caption{
Real space energy diagram illustrating  THz-
photon  ($f\gg \nu/2\pi$)  detection  in the superlattice:  DC  electric
field  $E_{SL}$  is   applied   to  the  $N$-period   semiconductor
superlattice with the miniband width $\Delta$ . Under the  action
of the dc field electrons perform Bloch oscillations with
the  spatial amplitude $\Delta/eE_{SL}$ . At critical dc electric voltage
(field)
$V_{SL}=V_P=N\hbar\nu/e$ ($eE_{SL}d = \hbar\nu$)  
electrons move against the dc electric  force
due  to  absorption of photons climbing up  the  Wannier-
Stark  ladder.  The  energy $2eV_P$  should  be  absorbed   from
external ac field in order to subtract one electron  from
the  external circuit. One half of this energy is  needed
for  the electron to overcome the potential barrier which
is  formed  by  the dc force, with the other  half  being
delivered  to  the lattice due to energy  dissipation.  A
quasi-classical description of the process  is  valid  if
$f\ll \Delta/\hbar$
when  allowed transitions between different Wannier-Stark
state exist.}
\label{fig3}
\end{figure}

     In our case the mechanism of the photon detection is
different  (see  Fig.\  \ref{fig3}).  Electrons  move  against  the
applied  dc electric force due to absorption of  photons.
At $V_{SL}=V_P$   the  responsivity is negative, indicating  that  one
electron is subtracted from the dc current flowing  through
the  superlattice when the energy  $2eV_P$ is absorbed  from  the
external ac field. One half of this energy is needed  for
the  electron to overcome the potential barrier which  is
formed by the dc force, with another half being delivered
to  the lattice due to energy dissipation. If the applied
dc  voltage is strong enough, i.e. $V_{SL}\gg V_P$, 
dissipation plays no
essential role in the superlattice responsivity. In  this
case the energy $eV_{SL}$ should be absorbed from the ac field  in
order to subtract one electron from the dc current simply
due to the energy conservation law.

\begin{figure}
\epsfxsize=120mm
\epsffile{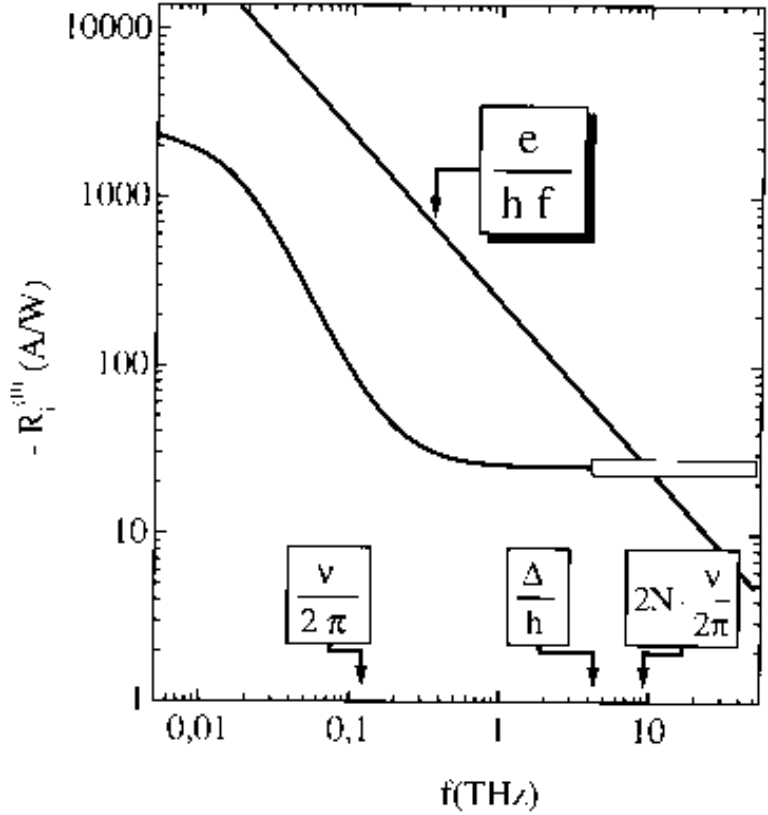}
\caption{
The calculated current responsivity 
$-R^{(0)}_i(f)$ of  the
ideally coupled superlattice with 
$d$ = 50 {\AA},
$N$ = 40, 
$E_P$  = 1  kV/cm,
$\Delta$= 20 meV  at sub-threshold voltage 
$V_{SL}=0.99 V_P$ (solid  curve),
and the quantum efficiency $e/hf$  (dashed curve) are plotted as
a  function of the frequency $f=\omega/2\pi$. 
At frequency 
$f_{QL}=2N\nu/2\pi$ the value of
the  responsivity  equals  the  quantum  efficiency.  The
calculation  of  the  responsivity is  valid  up  to  the
frequency  $f_\Delta=\Delta/h$ which  should  be smaller  than   
$f_{QL}=2N\nu/2\pi$ within  the
framework  of  the  employed model.  In  a  wide-miniband
superlattice
$\Delta\to 2N\times \hbar\nu$  the responsivity can approach  the  quantum
efficiency in the high-frequency limit.}
\label{fig4}
\end{figure}

     In  order  to  demonstrate typical frequency  scales
involved  in the problem we plot in Fig.\ \ref{fig4} the  frequency
dependence  of the responsivity 
$-R^{(0)}_i(\omega,V_{SL})$ for a superlattice  with
$d = 50${\AA},
$N=40$,
$L= 0.2\mu$m,
$E_P = 1$kV/cm,
$\Delta =20$meV,
$f_\nu = \nu/2\pi =0.12$THz
at  sub-threshold voltage 
$V_{SL}=0.99 V_P$ as a function of the frequency $f=\omega/2\pi$.  
This  example roughly corresponds to the  superlattice
samples  experimentally  investigated  in  Ref.  \onlinecite{ref33}.  
The
responsivity decreases in the region 
$f<f_\nu$ and, then, tends to
the  constant value determined by Eq. (\ref{(47)}). At  frequency
$f_{QL} = 2N\nu/2\pi$
the   value  of  the  responsivity  equals  the   quantum
efficiency $e/hf$.

     However, the semiclassical approach employed in  the
present  paper  is  restricted by the inequality 
$f\le f_\Delta = \Delta/h$.  This
inequality simply requires that there must be an  allowed
transition between different Wannier-Stark states in  the
superlattice   miniband  due  to  photon  absorption   as
illustrated  in  Fig.\  \ref{fig3}. It is important  to  note that  
$f_\Delta/f_{QL} = \ell_{FP}/L\le 1$
(where  $\ell_{FP} = v_0\tau=\Delta d/2\hbar\nu$ 
is  the  electron's  mean  free  path)  in   our
calculations and, therefore, the current responsivity  of
the   whole  sample  is  always  less  than  the  quantum
efficiency $e/hf$. In a wide-miniband superlattice  with
$\Delta \to 2N \hbar\nu$   (or,
equivalently, in a short superlattice sample with 
$L\to\ell_{FP}$ )  the
responsivity is approaching the quantum efficiency in the
high-frequency limit.

     As was mentioned above, an increase of the length $L$ of
a detector  may have an advantage in the reduction of the
parasitic capacitance $C\propto 1/L$, 
and, hence, in the increasing  of
the  roll-off frequency  $1/R_SC$ of the device. It is interesting
to  note that in case of  $N$ superconductor tunnel junctions
connected  in  series  the  resulting  responsivity   is
expected  to  be $e/N\hbar\omega$, implying that one photon  should  have
been  absorbed  in  each junction in  order  to  add  one
electron  to  the  dc  circuit\cite{ref18}.  In  the  case   of   a
superlattice samples consisting of  $N$ unit cells  (periods)
the resulting responsivity can be presented as 
$-e/2N\hbar\nu = (-e/N\hbar\omega)\times(\omega/2\nu)$. For  
$\omega>2\nu$ the
responsivity of the unit superlattice cell starts  to  be
larger than the quantum efficiency $e/\hbar\omega$.
 This conclusion  can
be  readily  understood  if  one  realises  that  due  to
absorption  of  one photon in the miniband electrons  can
`jump'  over  several superlattice periods as illustrated
in Fig.\ \ref{fig3}.

\section{RESULTS AND DISCUSSION}

     In  this  section we shall investigate  the  current
responsivity of the superlattice making use  of  the  
Eq.(\ref{(32)})  which  takes into account both the finite  matching
efficiency  between the antenna and the superlattice  and
the  parasitic  losses  in  the  superlattice  diode.  We
present  our  results  using  the  dimensional  value  of
$R_i(\omega,V_{SL})$
(given  in  units of Ampere / Watt) and also  introducing
the normalised responsivity
\begin{equation}
R_{iN}(\omega,V_{SL}) =
R_{i}(\omega,V_{SL})/(e/\hbar\omega)
\label{(48)}
\end{equation}
which permits us to compare directly the responsivity
$R_{i}(\omega,V_{SL})$  of
the  superlattice  with the quantum efficiency
$e/\hbar\omega$ .  In  our
calculations we use the typical geometrical dimensions of
the    superlattice   samples   investigated    in    the
experiments\cite{ref43,ref44,ref45} and also assume that the bow-tie antenna
impedance 
$Z_A=50 \Omega$  does not depend on THz-photon frequency\cite{ref33}.

\subsection{Excitation of the plasma-Bloch oscillations}

\begin{figure}
\epsfxsize=120mm
\epsffile{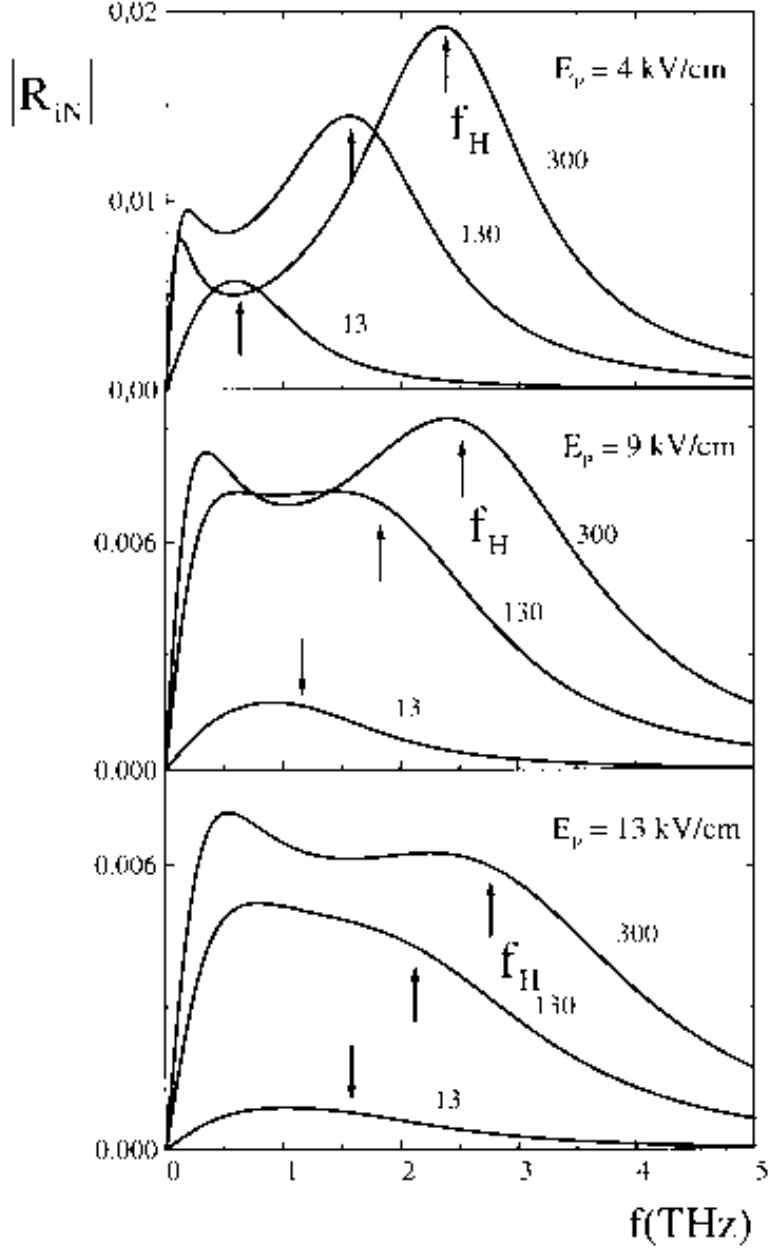}
\caption{The  frequency dependence of the  normalised
current  responsivity  
$|R_{iN}|=|R_i/(e/\hbar\omega)|$ 
of  the  superlattice  THz-photon
detector 
($a=2\mu$m, 
$L=0.5\mu$m, 
$R_S=10\Omega$,
$V_{SL} = 0.95 V_P$) is calculated for three values of  the
peak current density 
($j_P=$ 13, 30, and 300 kA / cm$^2$ ) and  for
three values of the peak electric field 
($E_P=$ 4, 9, and 13 kV/cm ).
The   relevant   positions  of  the  hybrid  plasma-Bloch
frequencies $f_H$  are  indicated for  each  curve  by  arrows
showing   characteristic  resonance  (high  peak  current
densities)  and  roll-off  (low peak  current  densities)
behaviour.}
\label{fig5}
\end{figure}

     For demonstration of the frequency dependence of the
superlattice  current responsivity in  the  THz-frequency
band we will focus on the GaAs/Ga$_{0.5}$Al$_{0.5}$As superlattices
specially   designed  to  operate  as   millimetre   wave
oscillators at room temperature. In Ref. \onlinecite{ref43} 
wide-miniband
superlattice    samples   with
$d=50${\AA},
$\Delta\simeq 113$meV,
$n\simeq 10^{17} {\rm cm}^{-3}$,    were   investigated
experimentally. They demonstrated a well-pronounced Esaki-
Tsu  negative differential conductance for
$E_{SL}\ge E_P \simeq 4$kV/cm  with the high
peak current of the order of 
$j_P \simeq 130$kA/cm$^2$. The measured value of  the
peak  current  is in a good agreement with  the  estimate
$j_P\simeq(80-160)$ kA/cm$^2$
for  
$n\simeq (1-2)\times 10^{17}$,  
$T=300$K  based  on Eq. (\ref{(12)}),  if  one  assumes  an
equilibrium   Boltzmann  distribution  for   the   charge
carriers.  From  the peak electric field and  current  we
find   the   scattering  and  plasma  frequencies  
$f_\nu \simeq 0.5$THz, $f_P = 2$THz,
respectively,   assuming   
$\epsilon_0 = 13$
for  the  average   dielectric
lattice   constant.   The  maximum  frequency   for   the
semiclassical approach to be valid for these samples is 
$f_\Delta \simeq 27$ THz.
     Figure  \ref{fig5}  shows  the frequency  dependence  of  the
normalised  current  responsivity  calculated  for  three
values  of  the peak current density in the superlattice,
i.e. $j_P=$ 13, 130,  and  300
kA/cm$^2$ and for three values  of  the  peak
electric field,
$E_P=$ 4,  9, and
13 kV/cm.  We  also use the typical values for the superlattice
length $L=0.5\mu$m   (superlattice  consists  of  100  periods),  and
assume $a=2\mu$m  for the superlattice mesa 
radius\cite{ref43,ref44,ref45}. We choose $R_S=10\Omega$
for  the  series  resistance of the device  in  the  THz-
frequency  band,  i.e.  the same value  as  for  resonant
tunnelling diodes having the same radius of mesas\cite{ref36}.  The
calculations are performed in the region of the  positive
differential conductance for dc bias voltage close to the
peak voltage ($V_{SL}=0.95 V_P$)

    For $E_P = 4$kV/cm  
($f_\nu\simeq 0.5$ THz) Fig.\ \ref{fig5} demonstrates well-pronounced resonant
behaviour of the normalised responsivity as a function of
frequency. The resonance frequency and the maximum  value
of  the  responsivity  rise if the peak  current  density
increases.  For 
$j_p = 300$ kA/cm$^2$ the normalised responsivity reaches  its
maximum value $-R_{iN}\simeq 0.02$
($-R_i \simeq 2$ A/W) at frequency $f\simeq 2.5$ THz .
    For higher values of the peak electric fields 
$E_P = 9$ kV/cm ($f_\nu\simeq 1.08$ THz) and
$E_P = 13$ kV/cm ($f_\nu\simeq 1.57$ THz)
the  resonance  line-widths  are  broadened  due  to
implicit  increase  of  the  scattering  frequencies.  In
particular,  for
$E_P = 13$ kV/cm,
$j_P=300$ kA/cm$^2$  the normalised responsivity  has  an
almost  constant  value  
$-R_{iN}\simeq 0.006$
($-R_i \simeq 0.6$ A/W) up  to $f\simeq 2.5$ THz   and,  then,  rapidly
decreases.

      The   frequency   behaviour   of   the   normalised
responsivity  originates from excitation of  the  plasma-
Bloch  oscillations in the superlattice. We  indicate  in
Fig.  \ref{fig5}  the  positions of the hybrid  frequencies
$f_H=|\omega^H_\pm|/2\pi$   with
arrows. For small peak electric fields (low values of the
scattering  frequencies) the hybrid frequency corresponds
to the maximum of the normalised responsivity. For higher
values of the peak field (higher values of the scattering
frequencies) it corresponds to the roll-off frequency  at
which the responsivity starts to decline.

\begin{figure}
\epsfxsize=120mm
\epsffile{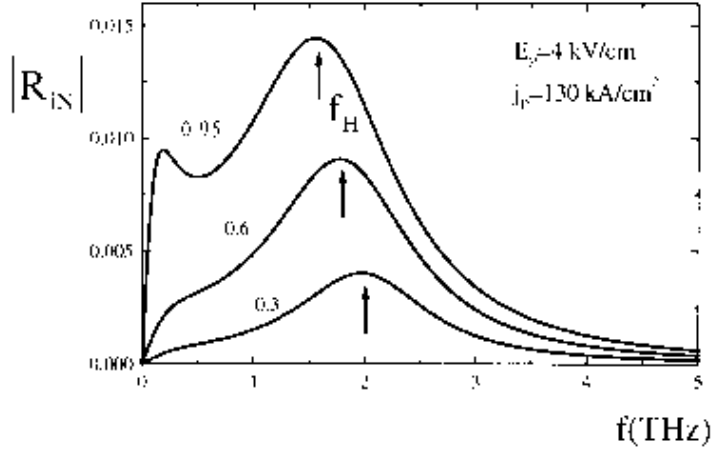}
\caption{The  frequency dependence of the  normalised
current  responsivity  
$|R_{iN}|=|R_i/(e/\hbar\omega)|$ 
of  the  superlattice  THz-photon
detector  
($a=2\mu$m,
$L=0.5\mu$m, 
$R_S=10 \Omega$, 
$E_P =$4kV/cm, 
$J_P=130$ kA/cm$^2$) is calculated for three  values  of
the  applied dc voltage 
$V_{SL}=$ 0.3, 0.6, and 0.95 $V_P$.  The  hybrid
plasma-Bloch frequencies $f_H$ (indicated for each dc  voltage
by  arrows)  are  found to be in a  good  agreement  with
positions of maximum values of the responsivity.}
\label{fig6}
\end{figure}

      Frequency   dependence  of  the  responsivity   for
different  applied dc fields , 
$V_{SL} = 0.3$, 0.6 and 0.95 $V_P$ is illustrated
in  Fig.\ \ref{fig6}  for
$j_P= 130$kA/cm$^2$. The same geometrical  dimensions  and
circuit  parameters of the superlattice device have  been
chosen  for  calculations as previously.  The  normalised
responsivity decreases with decreasing of the bias  field
tending  to zero at $V_{SL}\to 0$. On the other hand, the position  of
maximum   responsivity  shifts  to  lower   values   with
increasing of the bias field in full agreement  with  Eq.(\ref{(45)}).

\subsection{Optimised superlattice length}
      The  enhancement  of  the  normalised  responsivity
requires   an   optimum  matching   efficiency   of   the
superlattice  to the broad-band antenna and  minimisation
of  the  parasitic losses in the series  resistor.  These
requirements impose an optimum length of the superlattice
for each chosen frequency of the incoming THz-photons and
series resistance.

\begin{figure}
\epsfxsize=120mm
\epsffile{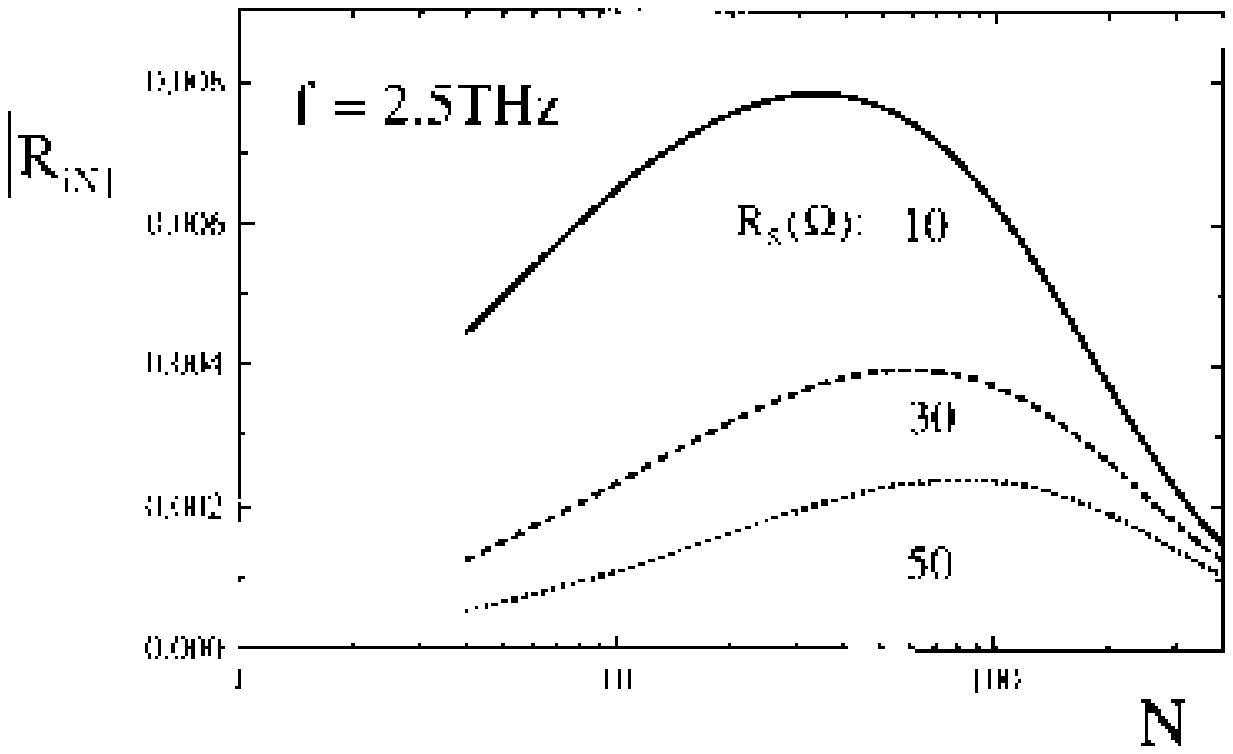}
\caption{The  dependence  of the  normalised  current
responsivity  
$|R_{iN}|=|R_i/(e/\hbar\omega)|$
of the superlattice THz-photon detector 
($a=2\mu$m,
$E_P=4$kV/cm,
$j_P=130$ kA/cm$^2$) is calculated at 
$f =2.5$ THz for three values of the
series  resistance  
($R_S$=10, 30, and 50 $\Omega$)  as  a  function  of
number of the superlattice periods $N$.}
\label{fig7}
\end{figure}

     We  show  in Fig.\ \ref{fig7} the dependence of the normalised
responsivity  on  the number of the superlattice  periods
for $f= 2.5$THz.We used for calculation 
$a=2\mu$m,
$j_P = 130$ kA/cm$^2$,
$V_{SL} = 0.95 V_P$,  and three values  of
the  series resistance $R_S = 10$, 30 and 50 $\Omega$. 
For all three  values
of the series resistance the responsivity displays a well
pronounced  maximum  for  the  optimum  number   of   the
superlattice  periods  $N=N_{\rm max}$. The value  of
$N_{\rm max}$   increases  with
increasing  of the series resistance (
$N_{\rm max}\simeq 40$ for $R_S =10 \Omega$,
$N_{\rm max}\simeq 60$ for $R_S =30 \Omega$  and
for 
$N_{\rm max}\simeq 90$ for $R_S =50 \Omega$). 
This result can be readily understood by recalling
that  a  larger  volume  of  the  superlattice  minimises
parasitic   losses  for  higher  values  of  the   series
resistance   because  of  reduction   of   the   sample's
capacitance.

\begin{figure}
\epsfxsize=120mm
\epsffile{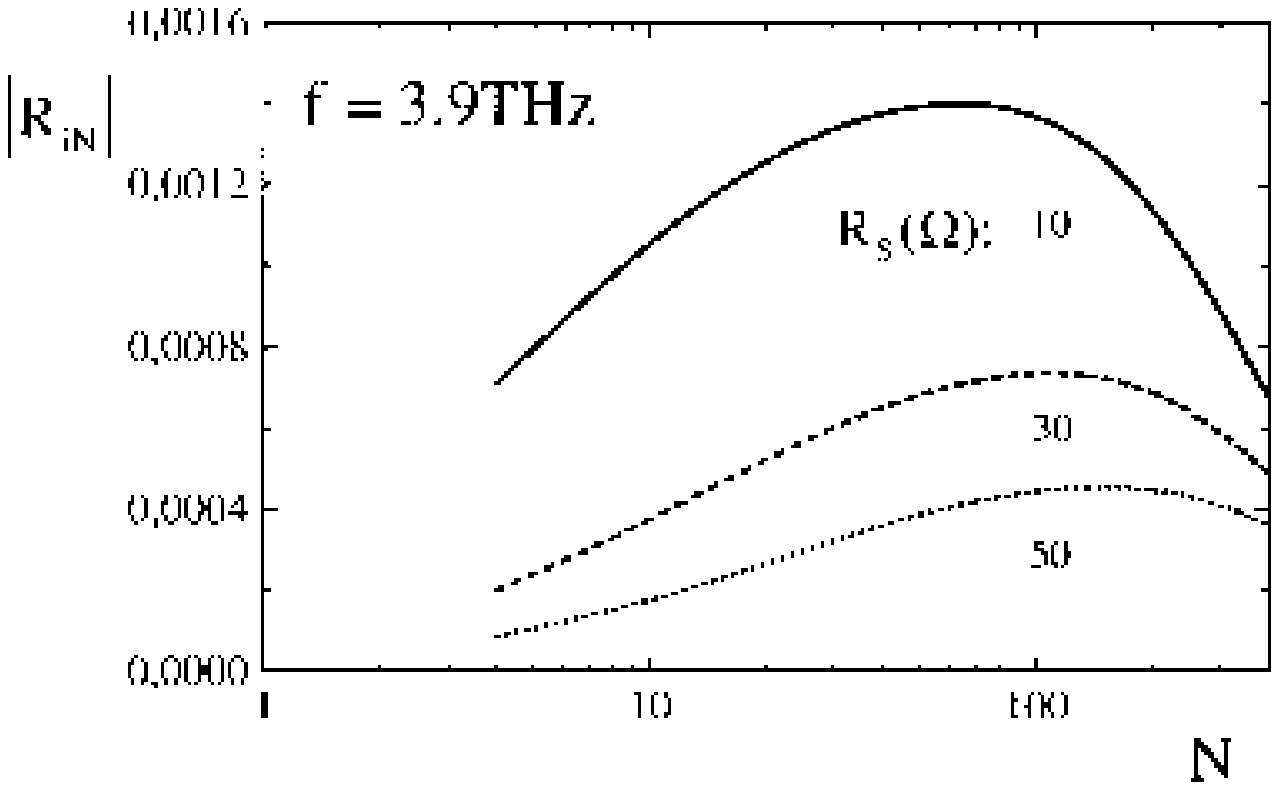}
\caption{
The  dependence  of  the  normalised  current
responsivity 
$|R_{iN}|=|R_i/(e/\hbar\omega)|$ 
of the superlattice THz-photon detector 
($a=2\mu$m,
$E_P=4$kV/cm,
$j_P=130$ kA/cm$^2$) is calculated at 
$f =3.9$ THz for three values of the
series  resistance  
($R_S$=10, 30, and 50 $\Omega$)  as  a  function  of
number of the superlattice periods $N$.}
\label{fig8}
\end{figure}

     For incoming photon higher frequencies the parasitic
losses   in  the  superlattice  device  play  even   more
important role leading to a further increase of
$N_{\rm max}$.  Figure \ref{fig8}
shows the dependence of the normalised responsivity on
the  number of the superlattice periods for
$f= 3.9$THz  and the same
superlattice parameters as in Fig.\ \ref{fig7}. 
In this case 
($N_{\rm max}\simeq 70$ for $R_S =10 \Omega$,
$N_{\rm max}\simeq 110$ for $R_S =30 \Omega$  and 
$N_{\rm max}\simeq 180$ for $R_S =50 \Omega$ ). 
We can conclude, therefore, that the
bulk   mechanism   of  the  superlattice   high-frequency
response provides important benefits for operation of the
superlattice detectors in the THz-frequency band.

\subsection{Responsivity dependence on bias voltage}

    Experimental studies of interaction of high-frequency
fields  with  the  miniband  electrons  in  semiconductor
superlattices  having  a  relatively  small  
($j_P\simeq 10$ kA/cm$^2$)   current
densities  were performed\cite{ref44,ref45} 
for frequencies
$f = 90$ GHz,  
$f = 450$ GHz  and
$f = 3.5$ THz. 
A field-induced reduction of the current through the
superlattice   was  reported,  with  a   different   bias
dependence  below  and  above a characteristic  frequency
(1THz). Distinct bias dependence of the responsivity  was
attributed  to quasistatic ($\omega\tau\ll 1$) and 
dynamic ($\omega\tau\gg 1$)  interaction
of  the  miniband electrons with ac fields at frequencies
below  and  above  1THz, respectively. The  intraminiband
relaxation   time  
($\tau \simeq 10^{-13}$ s)  was  estimated  from  experimental
data\cite{ref45}.

\begin{figure}
\epsfxsize=120mm
\epsffile{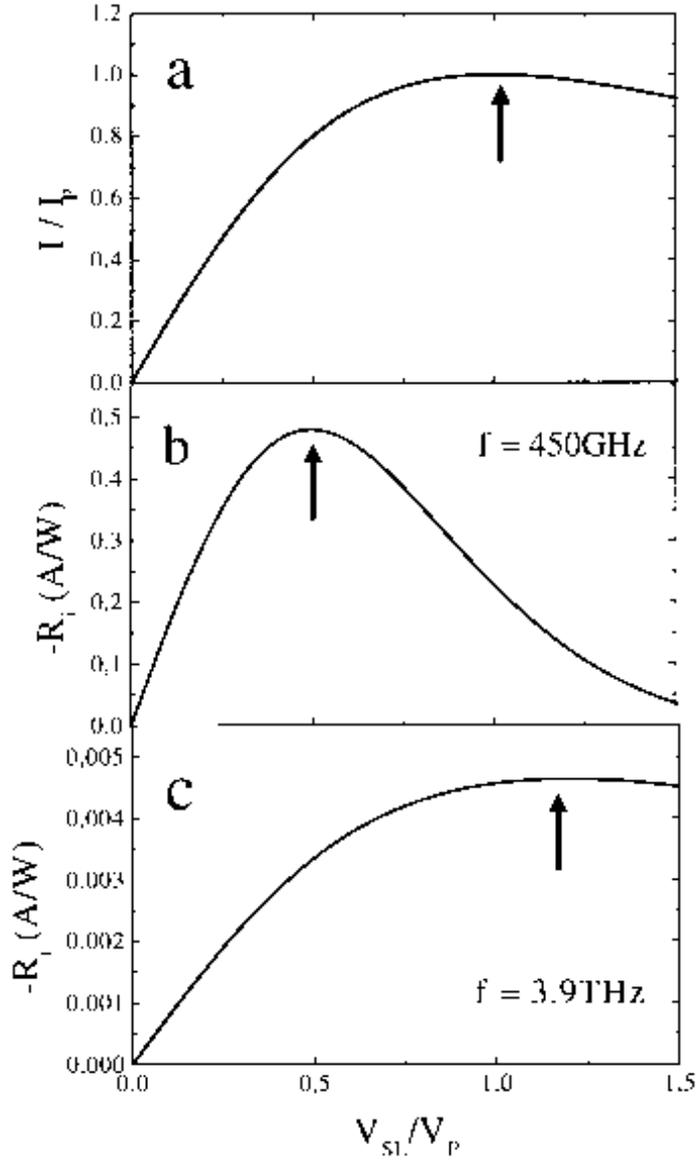}
\caption{(a) Normalised Esaki-Tsu $I-V$ characteristics
of  a  superlattice  THz-photon  detector  experimentally
investigated in Ref. 41 
($a=2\mu$m, 
$E_P=13$kV/cm, 
$j_P = 10$
kA/cm$^2$, 
$L=0.5\mu$m, 
$R_S= 20 \Omega$). Current responsivity
is  calculated for this detector for (b) 450-GHz and  (c)
3.9-THz  radiation  as a function of  the  normalised  dc
voltage  $V_{SL}/V_P$.}
\label{fig9}
\end{figure}

     In  Fig.\  \ref{fig9}  we plot the dc bias dependence  of  the
responsivity  at   
$f = 450$GHz and 
$f = 3.9$THz  for the superlattice  parameters
corresponding  to experiments\cite{ref45} 
($j_P\simeq 10$ kA/cm$^2$,
$E_P\simeq 13$ kV/cm,
$a\simeq 2.5\mu$m,
$L\simeq 0.5\mu$m,
$R_S\simeq 20\Omega$) together  with
the   dc  Esaki-Tsu  current-voltage  curve.  In  a  full
agreement  with  the observations\cite{ref45},  at
$f = 450$ GHz   responsivity
reaches  its maximum value at
$V_{SL}\simeq 0.5 V_P$  (roughly corresponding  to
the  position of the maximum of the second derivative  of
the  dc  current-voltage curve), while for  3.9 THz-radiation
the  position  of  the  maximum is shifted  to  the  peak
voltage $V_P$.  This can be readily understood if  one  takes
into  account  that  for $\omega\tau\ge 1$  the  model  of  the  classical
rectification  is  no  longer valid.  In  this  case  the
current  change under THz-photon irradiation is described
by
Eq.  (\ref{(22)}) giving the second derivative of the current  in
the  finite  difference  form taking  account  of  finite
photon  energies  .  At  high frequencies
$\omega\tau\to\infty$  this  equation
yields
$\Delta I^{SL}_{DC}(\omega,V_{SL})\propto I^{SL}_{DC}(V_{SL})$,   
i.e.  the  bias-field  dependence   of   the
responsivity  should  reproduce  the  dc  current-voltage
curve showing the maximum value at $V_{SL}\simeq V_P$.

\begin{figure}
\epsfxsize=120mm
\epsffile{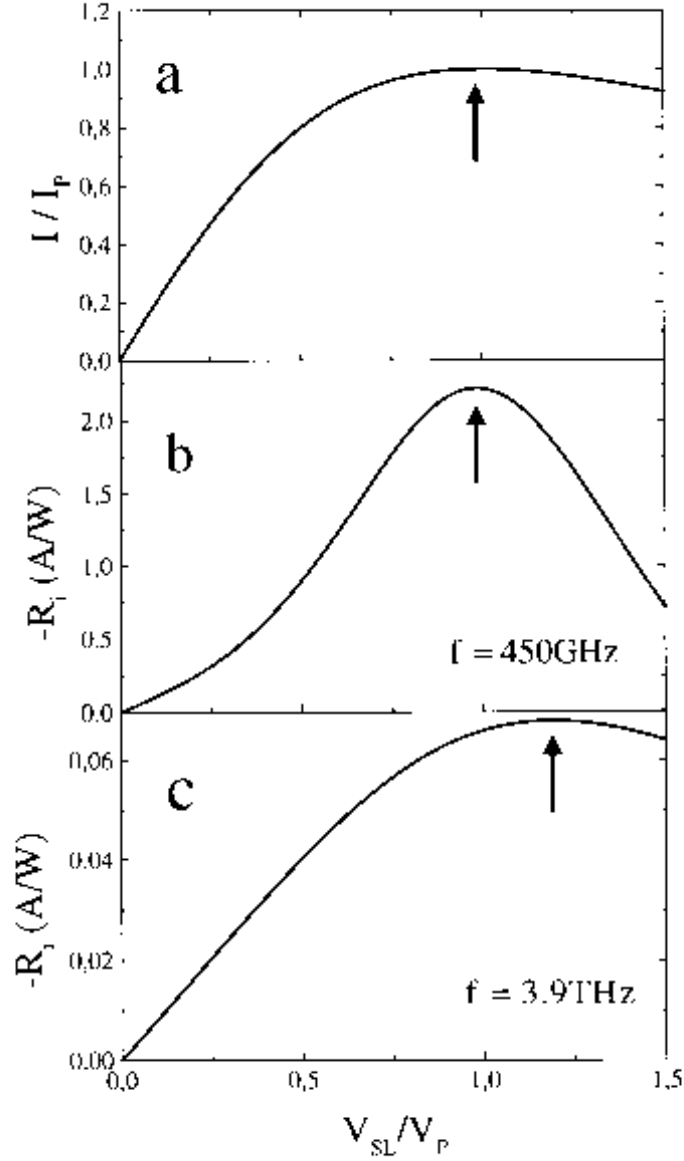}
\caption{(a) Normalised Esaki-Tsu $I-V$ characteristics
of   a   high  current  density  superlattice  THz-photon
detector  
($a=2\mu$m, 
$E_P=4$ kV/cm, 
$j_P= 130$ kA/cm$^2$, 
$L=0.5 \mu$m, 
$R_S=20\Omega$). Current responsivity is calculated
for  this  detector  for  (b)  450-GHz  and  (c)  3.9-THz
radiation as a function of the normalised dc voltage 
$V_{SL}/V_P$.}
\label{fig10}
\end{figure}

     It  is  important to note that with increasing  peak
current  densities this behaviour qualitatively  changes.
Figure  \ref{fig10} 
demonstrates the bias-field dependence of  the
responsivity  for  the same frequencies and  superlattice
parameters  as Fig.\ \ref{fig9} and for
$j_P\simeq 130 $kA/cm$^2$. First, we emphasise  that
the  responsivity  is considerably  higher  than  in  the
previous case both for 450 and 3.9THz radiation.  Second,
for  450 GHz-radiation the responsivity reaches  its  maximum
value  at 
$V_{SL}\simeq V_P$  and not at
$V_{SL}\simeq 0.5 V_P$  as previously. For the high  peak
current  densities  the superlattice impedance  variation
due   to  applied  dc  voltage  essentially  changes  the
coupling  efficiency  between antenna  and  superlattice.
This leads to a qualitatively different behaviour of  the
responsivity for low and high peak current densities: the
latter  can  manifest  in experiments  a  more  efficient
coupling of radiation into the superlattice.

\subsection{Optimised peak current density}

\begin{figure}
\epsfxsize=120mm
\epsffile{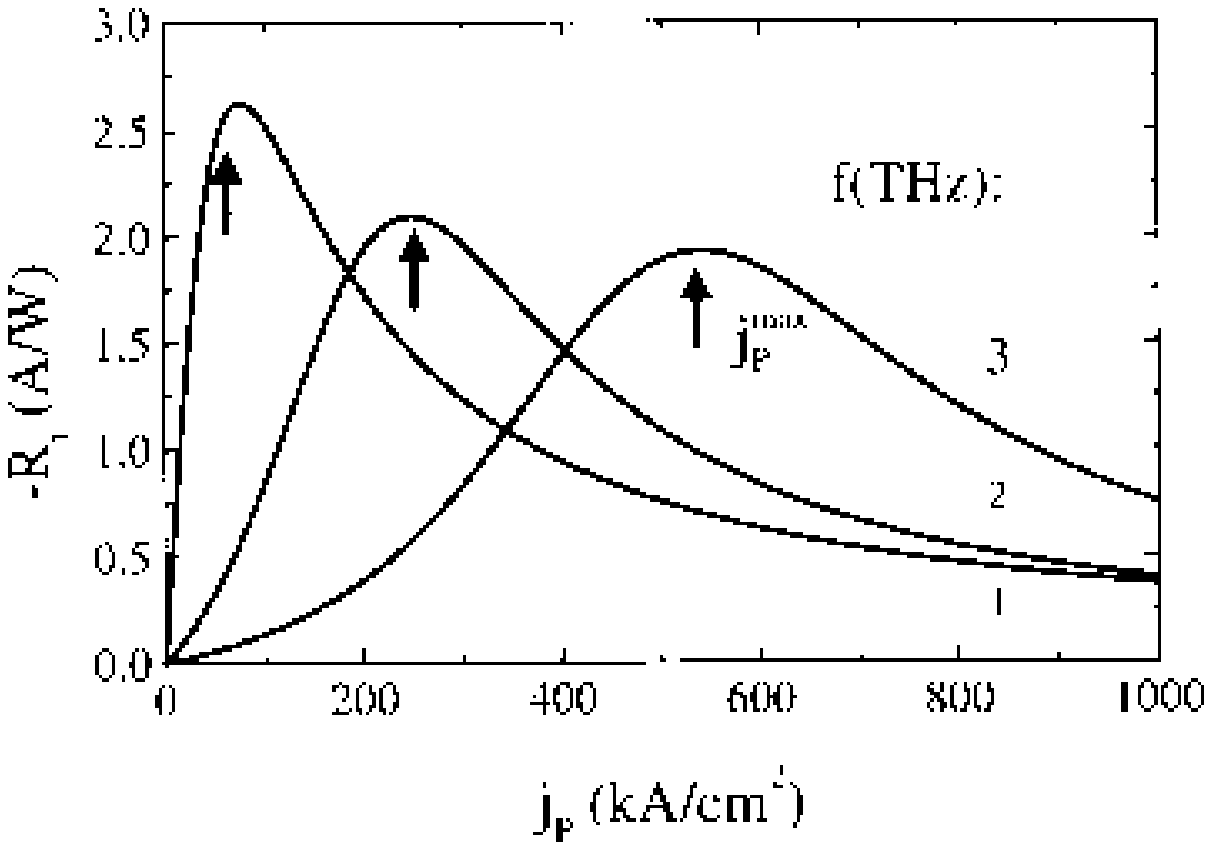}
\caption{
The current responsivity of the superlattice
THz-photon detector 
($a=2.5\mu$m, 
$E_P=4.5$kV/cm, 
$L=0.5\mu$m, 
$R_S=20\Omega$) is calculated as a  function
of  the  peak  current density for three  frequencies  of
radiation:  $f$ = 1, 2and 3 THz. The relevant positions  of
the peak current densities  
$j^{\rm max}_P$ calculated from criterion  of
excitation  of the plasma-Bloch oscillations
$f=f_H(j^{\rm max}_P)$   are  shown
for each frequency $f$  respectively.}
\label{fig11}
\end{figure}

      For  a  given  superlattice  geometry  the  current
responsivity of the superlattice can also be enhanced  by
choosing  an optimum value of the peak current densities.
This circumstance is illustrated in Fig.\ \ref{fig11} where we plot
the  responsivity  as  a function  of  the  peak  current
density  for
$V_{SL} = 0.99 V_P$,
$E_P\simeq 4$ kV/cm,  
$a\simeq 2.5\mu$m,  
$L\simeq 0.5 \mu$m,
$R_S\simeq 10\Omega$   and for three  frequencies  of
radiation, i.e.
$f=1 , 2$ and 3 THz. For different frequencies  the
responsivity  reaches its peaks at  different  values  of
$j^{\rm max}_P(f)$
which   increases  with  increasing  of   the   radiation
frequency.

    This behaviour can also be explained by excitation of
the plasma-Bloch oscillations in the superlattice by THz-
photons  if one takes into account the resonant condition
$f=f_H(j^{\rm max}_P)$,  
where the hybrid frequency 
$f_H=|\omega^H_\pm/2\pi|$ is given by Eq. (\ref{(45)})  (see
also Fig.\ \ref{fig2}). Calculated values
$j^{\rm max}_P(1$THz$)\simeq 60$kA/cm$^2$,
$j^{\rm max}_P(2$THz$)\simeq 220$kA/cm$^2$ and
$j^{\rm max}_P(3$THz$)\simeq 540$kA/cm$^2$ are shown in 
Fig.\ \ref{fig11}
by  arrows.  They  are  in  a  good  agreement  with  the
positions of the peaks found from calculations  based  on
the general Eq. (\ref{(32)}).

    We can conclude, therefore, that high current density
superlattices  should  be used in order  to  achieve  the
large responsivity values (2-3 A/W) in the 1-3 THz -frequency
band.  The  high  current densities can  be  obtained  by
choosing  wide-miniband and/or highly-doped samples.  For
example, according to Eq. (\ref{(14)}) in  superlattices with  
$\Delta\simeq$ 130 meV\cite{ref25}
the peak current density
$j_P\simeq$ 1000 kA/cm$^2$  can be reached for
$n\simeq 3\times 10^{17}$ cm$^{-3}$. 
In this case the equilibrium thermal excitation energy
($kT\simeq$ 26 meV) and the Fermi energy of degenerate electrons
($\epsilon_F\simeq$ 27 meV) are
considerably smaller than the miniband width. Hence,  Eq.
(\ref{(14)})  can  be  employed  for  the  peak  current  density
estimates both for room and low-temperature conditions.

     Finally  we would like to note that the measurements
of   the  responsivity  of  resonant-tunnelling  (double-
barrier) heterostructure diodes reported in Ref. 
[\onlinecite{ref36}]  were
carried  out at frequencies as high as 2.5 THz. The  reported
value  of the responsivity was of the order of several  $\mu$A/W.
As has been mentioned\cite{ref36}, this value is smaller by over an
order of magnitude than the responsivity of THz-frequency
Schottky  diodes in this frequency band. We believe  that
an optimised superlattice detector as discussed above may
have  some  advantages over these devices due  to  rather
high  expected  current responsivity in the THz-frequency
band.

\section{CONCLUSIONS}

      In  conclusion,  we  have  calculated  the  current
responsivity of a semiconductor superlattice detector  in
the   THz-frequency  band  based  on  equivalent  circuit
modelling.  Using  a  path  integral  solution   of   the
Boltzmann    equation    within    the    relaxation-time
approximation for the collision integral we have obtained
an  analytical  expression for the  responsivity,  taking
account  of (i) frequency dependent superlattice response
to  a  THz  field  caused  by dynamical  localisation  of
electrons,  (ii) a finite matching efficiency  between  a
detector  antenna  and the superlattice,  and  (iii)  the
presence  of parasitic losses in the device caused  by  a
series (contact) resistance.

     We  find  that the responsivity of the  superlattice
ideally  coupled  to the incident radiation  tends  to  a
finite  value  with increasing radiation frequency.  This
value is simply determined by the energy conservation law
governing  the THz-photons absorption in the presence  of
scattering processes.

      Excitation  of  plasma-Bloch  oscillations  in  the
superlattice (which are found to represent eigenmodes  of
the  system  in  the THz-frequency band) can  essentially
enhance  both the magnitude and the roll-off  frequencies
of the responsivity due to resonant coupling of radiation
into the superlattice. The excitation of the plasma-Bloch
oscillations  can  manifest itself  as  a  resonance-like
dependence  of the normalised responsivity  on  the  THz-
photon frequency and (or) as a specific dependence of the
responsivity  on  bias  fields for superlattices  showing
high peak current densities.

     Changes  in peak current density and in superlattice
length can effect drastically the coupling efficiency and
parasitic  losses  in  the  superlattice  device.   These
parameters are found to play an important role in the THz-
field  detector performance and need to be optimised  for
each   value  of  the  radiation  frequency  and   series
resistance. For higher frequencies the optimum length  of
the superlattice detector tends to increase demonstrating
a  benefit  of  a  bulk-type mechanism of the  THz-photon
detection.

     In  currently available superlattices possessing the
optimised length and the peak current the responsivity is
expected  to  be  as  high as 
$|R_i|\simeq$ (2-3) A/W  in the 1-3  THz-frequency
band.  These  values  of  the responsivity  range  up  to
several  percents of the quantum efficiency
$e/\hbar\omega$  of an  ideal
superconductor tunnel junction for this frequency  range.
The analysis of the current responsivity performed in the
present paper does not necessarily assume cooling of  the
superlattice: the   estimated  values   of   the   current
responsivity  of  superlattices can be expected  at  even
room temperature.

\begin{acknowledgements}

We would like to acknowledge fruitful discussions with K.
F.  Renk,  S.  J. Allen, N. J. M. Horing, A.  Wacker,  E.
Schomburg,  J.  Grenzer,  S. Winnerl,  E.  P.  Dodin,  A.
Zharov,   and  D.  G.  Pavel'ev.  A.  A.  I.   gratefully
acknowledges a guest professorship financed  through  the
NATO Science Fellowship Programme.
\end{acknowledgements}

\vfill\eject

\begin{references}

\bibitem[*]{AA}Permanent address: Institute for Physics of
Microstructures, Russian Academy of Science, 603600
Nizhny Novgorod, Russia.

\bibitem{ref1}
L. Esaki and R. Tsu, IBM J. Res. Dev. {\bf 14}, 61 (1970).


\bibitem{ref2} 
R. Tsu and L. Esaki, Appl. Phys. Lett. {\bf 19}, 246 (1971).

\bibitem{ref3}  
A.  A.  Ignatov and Yu. A. Romanov, Sov.  Phys.  Solid
State  {\bf 17},  2216 (1975); 
Phys. Status Solidi  B  {\bf 73},  327
(1976).

\bibitem{ref4} 
V. V. Pavlovich and E. M. Epstein, Sov. Phys. Semicond.
{\bf 10}, 1196 (1976).

\bibitem{ref5} 
 A.  A.  Ignatov  and Yu. A. Romanov, Radiophysics  and
Quantum Electronics (Consultants Bureau, N.Y., 1978) 
Vol. {\bf 21}, p. 90.

\bibitem{ref6}  
D. H. Dunlap and V. M. Kenkre, Phys. Lett. A {\bf 127},  438
(1988); Phys. Rev. B {\bf 37}, 6622 (1987).

\bibitem{ref7} 
M. Holthaus, Phys. Rev. Lett. {\bf 69}, 351 (1992).

\bibitem{ref8} 
D. H. Dunlap, V. Kovanis, R. V. Duncan, and J. Simmons,
Phys. Rev. B {\bf 48}, 7975 (1993).

\bibitem{ref9}  
X.-G. Zhao, X.-W. Zhang, S.-G. Chen, and W.-X.  Zhang,
Intern. Journ. of Mod. Phys. B {\bf 7}, 4215 (1993).

\bibitem{ref10} 
M. Wagner, Phys. Rev. B {\bf 49}, 16544 (1994).

\bibitem{ref11}  
X.-G. Zhao, R. Jahnke, Q. Niu, Phys. Lett. A {\bf 202}, 297
(1995).

\bibitem{ref12} 
J. Rotvig, A. P. Jauho, and H. Smith, Phys. Rev. Lett.
{\bf 74}, 1831 (1995).

\bibitem{ref13}  
A.  A. Ignatov, E. Schomburg, J. Grenzer, K. F. Renk,
and E. P. Dodin, Z. Phys. B {\bf 98}, 187 (1995).

\bibitem{ref14}  
J.  I{\~n}arrea and G. Platero, Europhys.  Lett.  {\bf 34},  
43 (1996).

\bibitem{ref15} K.  Johnsen, and A.P. Jauho, Phys. Rev.  B {\bf 57},  
8860 (1998).

\bibitem{ref16} M. B{\"u}tiker, Phys. Lett. A {\bf 96}, 365 (1983).

\bibitem{ref17} 
A. A. Ignatov, K. F. Renk, and E. P. Dodin, Phys. Rev.
Lett. {\bf 70}, 1996 (1993);
J. B. Xia, Phys. Rev. B {\bf 58}, 3565 (1998).

\bibitem{ref18} For a review, see J. R. Tucker and M. J. Feldman, Rev.
Mod. Phys. {\bf 57}, 1055 (1985).

\bibitem{ref19}  X.  L.  Lei, N. J. M. Horing, H. L. Cui,  and  K.  K.
Thornber, Appl. Phys. Lett. {\bf 65}, 2984 (1994); Z.  Phys.  B
{\bf 104}, 221 (1997).

\bibitem{ref20} A. N. Korotkov, D. V. Averin, and K. K. Liharev, Phys.
Rev. B {\bf 49}, 7548 (1994).

\bibitem{ref21} O. A. Tkachenko, D. G. Baksheyev, and V. A. Tkachenko,
J. Appl. Phys. {\bf 81}, 1771 (1997).

\bibitem{ref22} X. L. Lei, J. Appl. Phys. {\bf 82}, 718 (1997).

\bibitem{ref23} A. W. Ghosh, A. V. Kuznetsov, and J. W. Wilkins, Phys.
Rev. Lett. {\bf 79}, 3494 (1997).

\bibitem{ref24}  K.  N. Alekseev, E. H. Cannon, J. C. McKinney, F.  V.
Kusmartsev, and D. K. Cambell, Phys. Rev. Lett. {\bf 80},  2669
(1998).

\bibitem{ref25}  A.  Sibille, J. F. Palmier, H. Wang, and  F.  Mollot,
Phys. Rev. Lett. {\bf 64}, 52 (1990).

\bibitem{ref26}  H.  T.  Grahn, K. von Klitzing, K. Ploog, and  G.  H.
D"{\"o}ler, Phys. Rev. B 43, 12094 (1991).

\bibitem{ref27}  M.  Hadjazi, J. F. Palmier, A. Sibille, H.  Wang,  E.
Paris and F. Mollot, Electronics Lett. {\bf 29}, 648 (1993).

\bibitem{ref28}  E. Schomburg, K. Hofbeck, J. Grenser, T. Blomeier, A.
A. Ignatov, K.F. Renk, D.G. Pavel'ev, Yu. Koschurinov, V.
Ustinov,  A.  Zhukov, S. Ivanov, and P. S. Kop'ev,  Appl.
Phys. Lett. {\bf 71}, 401 (1997).

\bibitem{ref29}  C. Minot, N. Sahri, H. LePerson, J. F. Palmier, J. P.
Medus, J. C. Esnault, Superlattices \& Microstructures {\bf 23},
1323 (1998).

\bibitem{ref30} A. A. Ignatov, E. Schomburg, K. F. Renk, W. Schats, J.
F.  Palmiwer, and F. Mollot, Ann. Phys. (Leipzig) {\bf 3},  137
(1994).

\bibitem{ref31}  B. J. Keay, S. Zeuner, S. J. Allen, K. D. Maranowski,
A.  C.  Gossard, U. Bhattacharaya, and M. J. W.  Rodwell,
Phys. Rev. Lett. {\bf 75}, 4102 (1995).

\bibitem{ref32}  S. Zeuner, B. J. Keay, S. J. Allen, K. D. Maranowski,
A.  C.  Gossard, U. Bhattacharaya, and M. J. W.  Rodwell,
Phys.  Rev. B {\bf 53}, 1717 (1996); A. Wacker, A.P. Jauho,  S.
Zeuner, and S.J. Allen, Phys. Rev. B {\bf 56}, 13268 (1996).

\bibitem{ref33}  K. Unterrainer, B. J. Keay, M. C. Wanke, S. J. Allen,
D. Leopard, G. Medeiros-Ribeivo, U. Bhattacharaya, and M.
J.  W.  Rodwell, Phys. Rev. Lett. {\bf 76}, 2973 (1996);  Inst.
Phys.  Conf.  Ser.  (IOP Publishing Ltd,  1997)  No  155:
Chapter 10, p. 729.

\bibitem{ref34} T. Dekorsy, R. Ott, H. Kurz, and K"hler, Phys. Rev.  B
{\bf 51}, 17275 (1995).

\bibitem{ref35}  A.  A. Ignatov, E. Schomburg, J. Grenzer, S. Winnerl,
K.   F.   Renk   and   E.  P.  Dodin,   Superlattices   \&
Microstructures {\bf 22}, 15 (1997).

\bibitem{ref36}  T. C. L. G. Sollner, W. D. Goodhue, P. E. Tannenwald,
C.  D. Parker, and D. D. Peck, Appl. Phys. Lett. {\bf 43},  588
(1983).

\bibitem{ref37}  H.  C.  Torrey and C. A. Whitmer, Crystal  Rectifiers
(McGraw-Hill, New York, 1948), p. 336.

\bibitem{ref38}  A. A. Ignatov and V.I. Shashkin, Sov. Phys. JETP. {\bf 66},
526 (1987).

\bibitem{ref39}  A. A. Ignatov and V. I. Shashkin, Phys. Status Solidi
B {\bf 110}, K117 (1982); Phys. Lett. A {\bf 94}, 169 (1983).

\bibitem{ref40}  A.  Wacker and A. P. Jauho, Phys. Rev. Lett. {\bf 80},  369
(1998);  the  conclusions  on  impurity  scattering  have
recently been generalised to electron-phonon scattering,
A. Wacker et. al. (unpublished).

\bibitem{ref41}  R.  G. Chambers, Proc. Phys. Soc. (London) A {\bf 65},  458
(1952).

\bibitem{ref42}  A. Wacker, S. J. Allen, J. S. Scott, M. C. Wanke, and
A. P. Jauho, Phys. Status Solidi B {\bf 204}, 95 (1997).

\bibitem{ref43} E. Dutisseuil, A. Sibille, J. F. Palmier, F. Aristone,
F.  Mollot,  and V. Thietty-Mieg, Phys. Rev. B  {\bf 49},  5093
(1994).


\bibitem{ref44}  E.  Schomburg, A. A. Ignatov, J. Grenser, K. F. Renk,
D.  G.  Pavel'ev,  Yu. Koschurinov,  B.  Ja.  Melzer,  S.
Ivanov, S. Schaposchnikov, and P. S. Kop'ev, Appl.  Phys.
Lett. {\bf 68}, 1096 (1996).

\bibitem{ref45}  S. Winnerl, E. Schomburg, J. Grenser, H.-J. Regl,  A.
A. Ignatov, A.D. Semenov,
K.  F.  Renk,  D.  G. Pavel'ev, Yu. Koschurinov,  B.  Ja.
Melzer, V. Ustinov, S. Ivanov, S. Schaposchnikov, and  P.
S. Kop'ev, Phys. Rev. B {\bf 56}, 10 303 (1997).

\end{references}
\end{document}